\documentclass[12pt]{article}
\usepackage{graphicx, amsmath, amssymb, cite, setspace, color, relsize, bm, bbold, array, hyperref, dsfont, bbold, xcolor, cancel,soul}

\usepackage[top=1 in, bottom=1 in, left=.9 in, right=.9 in]{geometry}

\newcommand{\captionfonts}{\footnotesize} 
\makeatletter
\long\def\@makecaption#1#2{%
  \vskip\abovecaptionskip
  \sbox\@tempboxa{{\captionfonts #1: #2}}%
  \ifdim \wd\@tempboxa >\hsize
    {\captionfonts #1: #2\par}
  \else
    \hbox to\hsize{\hfil\box\@tempboxa\hfil}%
  \fi
  \vskip\belowcaptionskip}
\makeatother

\def\lsim{ \lower .75ex \hbox{$\sim$} \llap{\raise .27ex
\hbox{$<$}} }
\def\gsim{ \lower .75ex \hbox{$\sim$} \llap{\raise .27ex
\hbox{$>$}} }

\newcommand\T{\rule{0pt}{4.5ex}}       
\newcommand\B{\rule[-3.0ex]{0pt}{0pt}} 

\def\fnote#1#2{\begingroup\def\thefootnote{#1}\footnote{#2}
     \addtocounter{footnote}{-1}\endgroup}

\let\oldsqrt\sqrt
\def\sqrt{\mathpalette\DHLhksqrt}
\def\DHLhksqrt#1#2{%
\setbox0=\hbox{$#1\oldsqrt{#2\,}$}\dimen0=\ht0
\advance\dimen0-0.2\ht0
\setbox2=\hbox{\vrule height\ht0 depth -\dimen0}%
{\box0\lower0.4pt\box2}}

\begin{document}

\title{The decay of hot KK space}

\author{Adam~R.~Brown  \\
 \textit{\small{Physics Department, Stanford University, Stanford, CA 94305, USA}} }
\date{}
\maketitle
\fnote{}{\hspace{-.65cm}email: \tt{adambro@stanford.edu}}
\vspace{-.95cm}


\begin{abstract}
\noindent The non-perturbative instabilities of hot Kaluza-Klein spacetime are investigated. In addition to the known instability of hot  space (the nucleation of 4D black holes) and the known instability of KK space (the nucleation of bubbles of nothing by quantum tunneling), we find two new instabilities: the nucleation of 5D black holes, and the  nucleation of bubbles of nothing by thermal fluctuation. 
These four instabilities are controlled by two Euclidean instantons, with each instanton doing double duty via two inequivalent analytic continuations; thermodynamic instabilities of one are shown to be related to mechanical instabilities of the other. I also construct bubbles of nothing that are formed by a hybrid process involving both thermal fluctuation and quantum tunneling. 
There is an exact high-temperature/low-temperature duality that relates the nucleation of black holes to the nucleation of bubbles of nothing.
\end{abstract}

\thispagestyle{empty} 
\newpage

\section{Introduction} 
Empty Minkowski is 
 indefatigably
stable.  The positive energy theorem \cite{Schon:1979rg} guarantees that there is no state with the same asymptotics and same energy as the vacuum, and so nowhere for the vacuum to go.

There are two destabilizing elements we can add. 
One is a temperature. Hot space is unstable: Gross, Yaffe and Perry showed that it may nucleate a black hole that, if it is sufficiently large and therefore sufficiently cool, will grow forever \cite{Gross:1982cv}. The other is a compact extra dimension. Kaluza-Klein spacetime, even at zero temperature, is unstable: Witten showed that it may nucleate a `bubble of nothing' where the extra dimension pinches off and a hole appears in spacetime \cite{BoN}.

In this paper, I will combine these two complications. I will add to empty Minkowski both a compact extra dimension and  a nonzero temperature. Unsurprisingly, we will find that the instabilities both of \cite{Gross:1982cv} and of \cite{BoN} persist, and we will in addition find two more non-perturbative instabilities. We will find that as well as nucleating four-dimensional black holes (five-dimensional black strings that wrap the extra dimension) we may also nucleate five-dimensional black holes. And we will find that as well as  nucleating bubbles of nothing by quantum tunneling, we may also nucleate bubbles of nothing by thermal fluctuation.

\begin{figure}[h!] 
   \centering
   \includegraphics[width=4.5in]{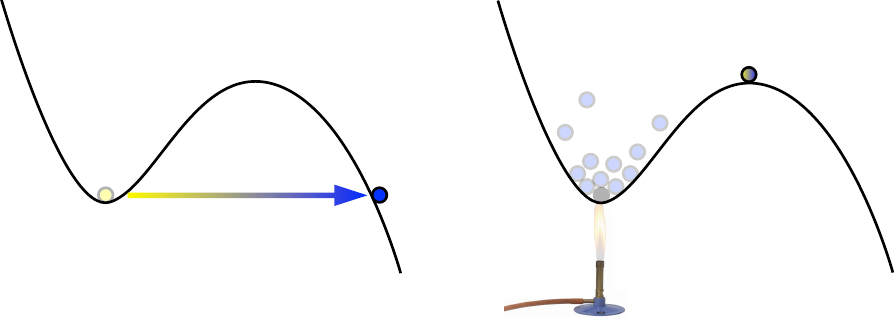} 
   \caption{In one-dimensional quantum mechanics there are two pure strategies for traversing obstacles---quantum tunneling through the barrier (rate $\sim \exp[-\frac{1}{\hbar}]$), or thermal fluctuation over the barrier (rate $\sim \exp[-\frac{1}{T}]$). Often  the dominant process will be a mixed strategy that fluctuates partway up the barrier then tunnels through the rest. In this paper we will see that the same is true for the nucleation of bubbles of nothing. The traditional Witten bubble of nothing results from pure quantum tunneling; we will construct bubble-of-nothing instantons that proceed by pure thermal fluctuation, as well as others that are part-thermal-part-quantum.} 
   \label{fig-twostrategies}
\end{figure}

The degrees of freedom of our theory are those of five-dimensional Einstein gravity, so the nonzero temperature fills spacetime with a thermal gas of gravitons. This thermal gas, as well as introducing the non-perturbative instabilities that will be the subject of this paper, also introduces perturbative instabilities such as the Jeans instability \cite{Jeans} famously responsible for the existence of the Earth. 
 Given the presence of a perturbative  instability, the question of the existence of a non-perturbative instability, and therefore the results of \cite{Gross:1982cv} and of this paper, may seem moot. However, as we will see in Appendix~\ref{appendix:perturbative}, the perturbative instabilities occur on a wavelength that is much  longer than the characteristic size of the non-perturbative processes. This means that with suitable IR boundary conditions \cite{Hawking:1976de,Hawking:1982dh} we can kill the perturbative instabilities entirely and isolate the non-perturbative instabilities as the only possible decays. This is possible so long as we keep the temperature and the KK scale safely sub-Planckian. This same limit will justify using the semiclassical and analogous `semicold' approximations, as well as allowing us to neglect the gravitational backreaction of the uncondensed radiation while calculating our Euclidean instantons.

Because we have both a nonzero temperature and a Kaluza-Klein extra dimension, the Euclidean instantons that mediate our decays will have two compact  directions---they are vacuum solutions to Einstein's equations with asymptotic geometry $R^3 \times S^1 \times S^1$. One of the $S^1$s is the extra dimension and has asymptotic circumference $L$; the other is the thermal circle and has asymptotic circumference $\beta \equiv \hbar/T$. Neither of the instantons we will consider are symmetric with respect to swapping the $S^1$s, which means that  each instanton describes two different decays, depending on which of the compact Euclidean directions is taken to represent the thermal circle. Even though there are four non-perturbative decay processes, we will find that there are only two distinct instantons, and that each instanton does double work.

For example,  the Euclidean black string wraps one of the $S^1$s but not the other. If we take the $S^1$ it wraps to be the extra dimension, then the instanton describes the nucleation of a black string; if we take the $S^1$ it wraps to be the thermal circle, then the instanton describes the thermal nucleation of a bubble of nothing. The Euclidean black hole does a similar double duty, describing both the nucleation of a five-dimensional black hole and the quantum nucleation of a bubble of nothing.  

Since relabeling the axes doesn't change the Euclidean action, it doesn't change the decay rate. Thus the rate to nucleate a black hole when the temperature is $T$  and the extra dimension has size $L$ is the same as the rate to nucleate a bubble of nothing when the temperature is $\hbar L^{-1}$ and the extra dimension has size $\hbar T^{-1}$. Since this relates large values of $LT$ to small values of $LT$, this is a high-temperature/low-temperature duality. 

(Afficionados of complex structure moduli will know that the geometry of a $T^2$ is  characterized not only by the length of its sides but also by an angle that deforms the rectangle into a parallelogram. Thermodynamically, this corresponds to turning on a Kaluza-Klein chemical potential. In this paper we will set the chemical potential to zero, but I will return to this complication in a forthcoming work \cite{me:chemicalpotential}, and we will see how the high-temperature/low-temperature duality fits into a larger SL$(2,\mathbb{Z})$ symmetry.) 

\pagebreak

The $\beta \leftrightarrow L$ duality is visible in the rate for our four decays, which we will see are given by 
 \begin{center}
\begin{tabular}{c|c|c} 
decay  &   instanton & rate  \B \\
  \hline   \hline
`quantum' bubble of nothing  &  black hole &  $ \exp\Bigl[ - \frac{1}{32 \pi} \frac{L^3}{\ell_{5}^3} \bigl( 1 - \frac{L^2}{16 \beta^2 } +   \frac{L^4 }{128 \beta^4 }  +  \ldots  \bigl) \Bigl]$ \T\B \\
\hline `thermal' bubble of nothing  & black string  &   $ \exp\Bigl[ - \frac{1}{16 \pi} \frac{\beta L^2}{\ell_{5}^3}\Bigl]$  \T\B \\
\hline 4D black hole   & black string &  $ \exp\Bigl[ - \frac{1}{16 \pi} \frac{\beta^2 L}{\ell_{5}^3}\Bigl]$  \T\B \\
\hline 5D black hole   & black hole  & $ \exp\Bigl[ - \frac{1}{32 \pi} \frac{\beta^3}{\ell_{5}^3} \bigl( 1 - \frac{\beta^2}{16 L^2} +   \frac{\beta^4 }{128 L^4 }  + \ldots  \bigl) \Bigl]$  \T \B
\end{tabular}
\end{center}
\vspace{-9mm}
\begin{figure}[htbp] 
   \caption{The four non-perturbative decays of hot KK space; the temperature is $\hbar/\beta$, the extra dimension has circumference $L$, and  $G_5 \equiv \ell_5^3/\hbar =  L G_4 \equiv  L \ell_4^2/\hbar$. Expressed in 4D coordinates the rate to nucleate a 4D black hole is independent of $L$.  There is a $\beta \leftrightarrow L$ duality that permutes the decays.}
   \label{fig:BHvsBScomparison}
\end{figure}
The rate to nucleate a 4D black hole agrees with that in \cite{Gross:1982cv}; the rate to nucleate a quantum bubble of nothing corrects by a factor of 2 the exponent of \cite{BoN}. 

The exponents for the black-string-instanton decays given in this chart are exact. But the exponent for the quantum bubble of nothing is only exact when $T=0$: for $T>0$ decay is faster because of a thermal assist (from fluctuating partway up the barrier). Similarly, the exponent for the 5D black hole is only exact when $L=\infty$: for $L<\infty$ decay is faster because of the nucleated black hole's gravitational attraction to its images. Though the interpretations differ, the duality guarantees the speed-up is the same. To calculate the exact speed-up involves a complicated numerical solution of coupled PDEs, mercifully this is a heroic calculation that has already been done  in a different context by someone else \cite{Headrick:2009pv,Sorkin:2003ka} and whose results I will first steal and then double analytically continue. The final results are plotted in Fig.~\ref{fig-altogether}. \\

The duality maps thermodynamic instabilities to mechanical instabilities. For example, both the black hole and the bubble of nothing need to be large to persist. The reasons for this are on the face of it quite different. The black hole  must be large in order to be cool---a black hole that is too small will have a Hawking temperature greater than that of the ambient gas and will evaporate down and be reclaimed by the heat bath. The bubble of nothing must be large for the same reason that all Coleman-De Luccia bubbles (of which the bubble of nothing is one limit \cite{BlancoPillado:2010df,Brown:2011gt}) must be large---so that 
the surface tension trying to contract the bubble loses to the pressure differential trying to expand it. These two seemingly quite different conditions---one mechanical, the other quantum mechanical---are 
dual to one another. 

Similarly, the temperature below which thermally fluctuating all the way to the top stops being a locally optimal way to make a bubble of nothing (what in de Sitter space is known as the Steinhardt-Jensen transition \cite{Jensen:1983ac,Hackworth:2004xb} of the Hawking-Moss instanton \cite{Hawking:1981fz}) is dual to the onset of the Gregory-Laflamme instability of a black string \cite{Gregory:1993vy}. In general, the dual bubble-of-nothing viewpoint will provide an enlightening alternative perspective on the properties of higher-dimensional black holes and black strings. But first let's return to quantum mechanics.

\subsection{Review of thermal and quantum decay}
\label{subsec:thermalandquantum}

In this subsection I will pedantically review the theory of thermally-assisted decay in one-dimensional quantum mechanics and highlight some notable phenomena that will persist for the bubble of nothing. 

\begin{figure}[h!] 
         \centering
   \includegraphics[width=3.7in]{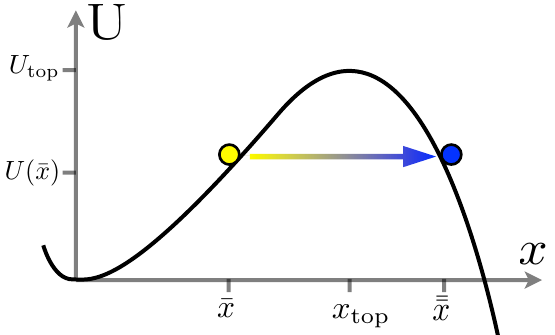} 
   \caption{To traverse the barrier, a particle may thermally fluctuate to $\bar{x}$, and then quantum mechanically tunnel to $\bar{\bar x}$. Since the quantum part of the process conserves energy, $U(\bar{x}) = U(\bar{\bar{x}})$. In the semiclassical description, after tunneling the particle appears at $\bar{\bar{x}}$ at rest, before classically rolling out to large ${x}$.} 
         \label{fig-halfwayupthenacross}
\end{figure}

If $\hbar = T = 0$, a ball in a local minimum of a potential is stuck. It can be unstuck by introducing either quantum mechanics or thermodynamics. Quantum mechanically ($\hbar > 0$) it may tunnel through the barrier; thermally ($T > 0$) it may fluctuate to the top of the barrier. With both quantum mechanics and thermal physics in play, it may also adopt a hybrid strategy 
the semiclassical description of which has three distinct steps: first thermally fluctuate to $\bar{x}$, then quantum tunnel to $\bar{\bar{x}}$, then classically roll from rest at $\bar{\bar{x}}$ out towards large $x$. The rate is 
\begin{eqnarray}
\Gamma & \sim & \exp \hspace{1pt} \left[ - \frac{1}{T} U(\bar{x}) \ \ \ -   \ \ \ \frac{2}{\hbar} \int_{\bar{x}}^{\bar{\bar{x}}} dx \sqrt{2[U(x) - U(\bar{x})]} \right] \label{eq:GammaMixedQM} \\
& \sim & \exp \left[- \frac{1}{T}   \vcenter{\hbox{\includegraphics[width=1in]{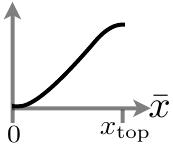}}}  \ \  \ \ \ - \ \     \ \, \vcenter{\hbox{\includegraphics[width=1in]{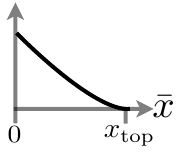}}} \  \right].
\label{eq:GammaMixedQMfigures}
\end{eqnarray}
The first term, the Boltzmann suppression factor, wants $\bar{x}$ to be as low as possible; the second term, the  WKB quantum tunneling factor, wants $\bar{x}$ to be as high as possible. 
There is an optimal tradeoff \cite{Affleck:1980ac}
given by 
\begin{equation}
\frac{\partial \Gamma}{\partial \bar{x}} = 0 \ \ \ \rightarrow \ \ \ \frac{1}{T} = \frac{2}{\hbar} \int_{\bar{x}}^{\bar{\bar{x}}} \frac{dx}{\sqrt{2[U(x) - U(\bar{x})]}} . \label{eq:periodofoptimum}
\end{equation}  

An alternative way to derive the same decay rate is to look for `instantons'. These are Euclidean solutions with periodicity $\beta$ and one negative mode that extremize the Euclidean action 
\begin{equation}
I_E = \int_0^{\beta} d \tau \left( \frac{1}{2} \Bigl( \frac{d{x}}{d \tau}\Bigl)^2 + U(x) \right) ;
\end{equation}
the decay rate is then $\exp[ - I_E/\hbar]$. To extremize the action, the instanton must obey $\frac{1}{2} (\frac{dx}{d \tau})^2 = U(x) - U(\bar{x})$ for some $\bar{x}$.
Only some values of $\bar{x}$ will give rise to trajectories with the right periodicity, but by plugging the instanton's  equation of motion into Eq.~\ref{eq:periodofoptimum} we see that those are exactly the values of $\bar{x}$ that extremize the decay rate. 
In this way of looking at things, by insisting on  periodicity $\hbar/T$ we automatically choose the optimum value of $\bar{x}$.

\begin{figure}[h!] 
   \centering
   \includegraphics[width=5in]{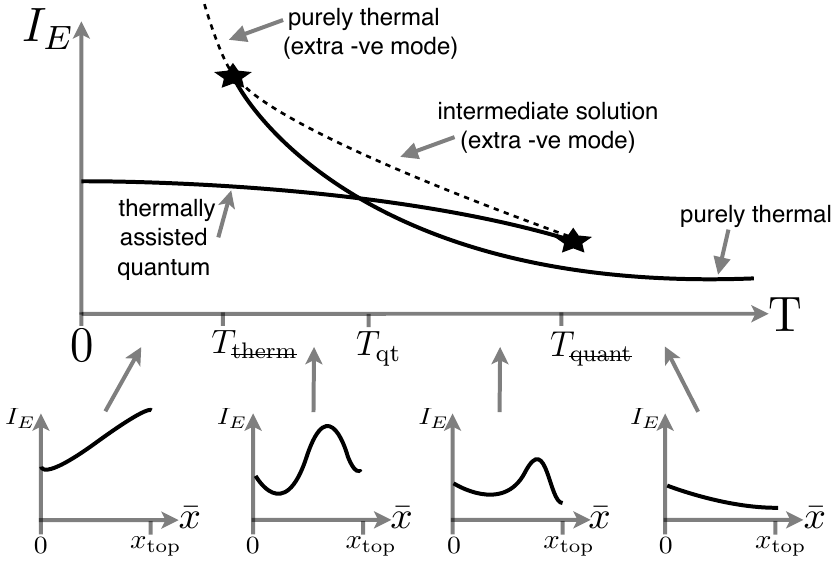} 
   \caption{The Euclidean action of the instantons that traverse the potential of Fig.~\ref{fig-halfwayupthenacross}, as a function of the temperature $T$. The tunneling rate is $\exp[- I_E/\hbar]$, so smaller $I_E$ means faster tunneling. 
Because of the thermal assist the quantum tunneling instanton has an action that falls with $T$. The intermediate solution chooses the worst possible value of $\bar{x}$---it gives the slowest case---and consequently has an extra negative mode.} 
   \label{fig-thermalandquantumQM}
\end{figure}
Or, rather, by insisting on periodicity $\hbar/T$ we automatically choose a local extremum value of $\bar{x}$, which need not be the global or even local optimum. There may be more than one extremum, which means more than one solution with Euclidean periodicity $\beta$; how many there are depends on the potential and depends on the temperature. Figure \ref{fig-thermalandquantumQM} shows the phase diagram of instantons that traverse the potential of Fig.~\ref{fig-halfwayupthenacross}. There are four regimes.  

\begin{enumerate}
\item $T > T_{\textrm{\st{quant}}}$: thermal only.

At high enough temperatures, the fastest way across the barrier is to go straight to the top, with no quantum tunneling at all.  The corresponding `pure thermal' instanton has $x(\tau) = x_{\textrm{top}}$ for all $\tau$: the instanton has a U(1) symmetry around the thermal circle. The action of this instanton recovers the classical Boltzmann rate, $I_E =  \beta U_{\textrm{top}}$. 
\item $T_{\textrm{\st{quant}}} > T > T_\textrm{qt}$: thermal beats thermally-assisted quantum.

For $T<T_{\textrm{\st{quant}}}$ there is another locally optimal way across the barrier. The corresponding `thermally-assisted quantum' instanton goes from $\bar{x}$ at $\tau =0$ to $\bar{\bar{x}}$ at $\tau = \beta/2$ (and then back to $\bar{x}$ at $\tau = \beta$ as required by periodicity) with a path given by $\frac{1}{2} \dot{x}^2 = U(x) - U(\bar{x})$: the instanton has only a Z$_2$ symmetry in the direction of  the  thermal circle.

The two (locally optimal) instantons are separated by a (locally pessimal) intermediate solution with an extra negative mode.

\item $T_{\textrm{qt}} > T > T_{\textrm{\st{therm}}}$: thermally-assisted quantum beats thermal.

At $T=T_{\textrm{qt}}$ the instantons exchange dominance in a first-order transition\footnote{There are other  possibilities for the phase diagram. By deforming the potential we can  either conjure additional instantons (with corresponding extra-negative-mode solutions), or get rid of the intermediate solution entirely by making the transition second-order \cite{Brown:2011um,Hackworth:2004xb}. However, the scheme described above is a typical pattern, and will also turn out to be that exhibited by bubbles of nothing in  hot  5D KK space.}.
\item $T_{\textrm{\st{therm}}} > T$: thermally-assisted quantum only. Pure thermal has extra negative mode(s).

For temperatures below
\begin{equation}
T_{\textrm{\st{therm}}} = \frac{\hbar \, \omega_{\textrm{top}}}{2 \pi}      \equiv \frac{\hbar}{2 \pi}    \sqrt{- \frac{V''_{\textrm{top}}}{m}},
\end{equation}
the pure thermal solution is not even a locally optimal way across the barrier and has extra negative modes.

As $T$ falls, the optimal $\bar{x}$ falls with it: the  instanton becomes less thermal and more quantum. But  the Boltzmann factor is quadratic in $\bar{x}$ near zero, whereas the WKB factor is linear, so for any nonzero temperature the optimal process has $\bar{x}$ strictly positive. 
\end{enumerate}

\begin{figure}[h!] 
   \centering
   \includegraphics[width=\textwidth]{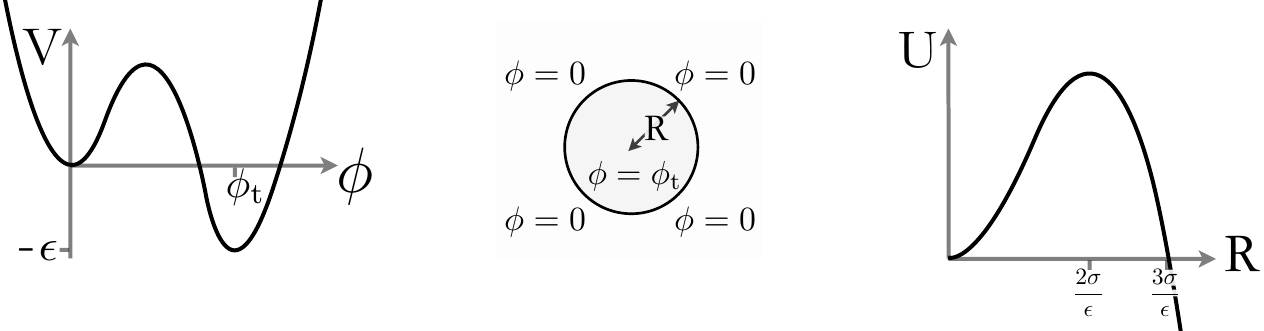} 
   \caption{Left: the potential energy density $V(\phi)$ of an example field. Center: a bubble of $\phi_\textrm{t}$ separated by a thin wall from the background $\phi = 0$. Right: the energy $U[R]$ of a static bubble of radius $R$ in flat spacetime.} 
   \label{fig-VvsU}
\end{figure}

These results generalize straightforwardly to thermally-assisted decay in quantum field theory. Quantum field theory is the quantum mechanics not of the disembodied field value $\phi$, but of the three-dimensional field configuration $\phi(\vec{x})$, and the correct analogue of $U(x)$ is not $V(\phi)$ but $U[\phi(\vec{x})] = \int d^3 x [ \frac{1}{2} (\vec{\nabla} \phi)^2 + V(\phi) ]$. The field does not homogeneously traverse $V(\phi)$ (the corresponding tunneling exponent would scale like the spatial volume), instead $\phi(\vec{x})$ traverses $U[\phi(\vec{x})]$ by inhomogenous  bubble nucleation. As depicted in Fig.~\ref{fig-VvsU}, in the thin-wall regime the pertinent family of field configurations have a `bubble' of lower vacuum, $\phi(r<R) = \phi_\textrm{t}$, separated by an interpolating wall from the ambient $\phi(r>R) = 0$. Neglecting gravity, and defining the wall tension as $\sigma \equiv \int_0^{\phi_\textrm{t}} d \phi \sqrt{2 V(\phi)}$, the energy of a static bubble of radius $R$ is 
\begin{equation}
U(R) = 4 \pi \sigma R^2 - \frac{4 \pi}{3} {\epsilon}  R^3.
\end{equation}
This function goes from zero at  $R=0$ to minus infinity as $R \rightarrow \infty$, but to get from one to the other---to go from no bubble to a huge bubble---involves traversing a barrier.

\begin{figure}[h!] 
   \centering
   \includegraphics[width=5in]{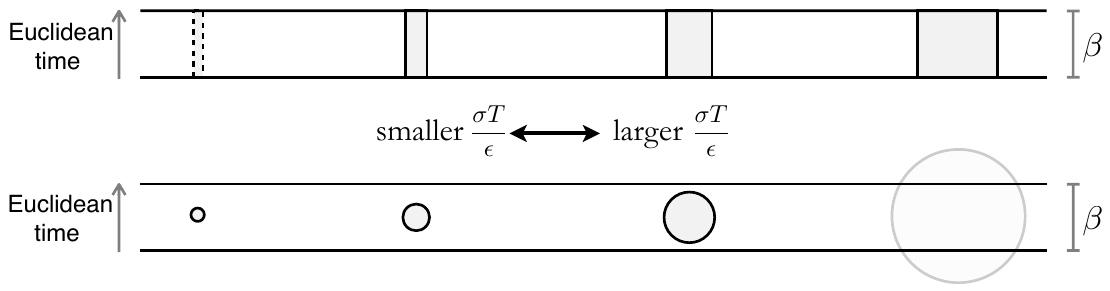} 
   \caption{Thermal (top) and thermally-assisted quantum (bottom) flat-spacetime QFT instantons as a function of $\sigma T/\epsilon$; the larger $\sigma T /\epsilon$, the larger the bubble needs to be relative to the thermal circle. (Since we are by assumption in fixed flat spacetime, the periodicity of the thermal circle is fixed at $\beta$; if we turned on gravitational backreaction the thermal circle would shrink near the center of the bubbles due to gravitational blueshifting, \`a la Tolman-Oppenheimer-Volkoff \cite{Oppenheimer:1939ne}.) The thermal instantons have a U(1) symmetry around the thermal circle; the quantum instantons have only a Z$_2$ symmetry. The field first thermally fluctuates to $\phi(\vec{x},\tau=0)$ [the horizontal slice represented by the black lines---this is the analogue of $\bar{x}$], then quantum tunnels to $\phi(\vec{x},\tau = \beta/2)$ [the horizontal slice through the center of the bubbles---the analogue of $\bar{\bar{x}}$].}
   \label{fig-thermalandquantumQFT}
\end{figure}

As before, the barrier may be traversed in two ways: there is the thermal way \cite{Linde:1981zj}, and there is the quantum way \cite{Coleman:1977py}. The thermal instanton has a U(1) symmetry around the thermal circle and $U'(R) = 0 \rightarrow R = 2 \sigma/\epsilon$; the quantum instanton has only a Z$_2$ symmetry around the thermal circle and $U(R) = 0 \rightarrow R = 3 \sigma/\epsilon$. These instantons are shown in Fig.~\ref{fig-thermalandquantumQFT} and  have a phase diagram analogous\footnote{One artifact of being in  the thin-wall regime is that the thermal assist of the quantum instanton is  tiny.} to Fig.~\ref{fig-thermalandquantumQM}.

\newpage

After nucleation, a bubble bears the imprint of how it was made: the thermal-made bubble is born smaller than the quantum-made bubble ($R = \frac{2 \sigma}{\epsilon}$ \emph{vs.} $R = \frac{3 \sigma}{\epsilon}$); the thermal bubble has positive energy whereas the quantum bubble has zero energy; and the thermal bubble is  in (unstable) static equilibrium whereas the quantum bubble, while born  at rest, immediately accelerates outwards. 
We'll see these features repeated for  bubbles of nothing.

Finally, let's connect this analysis with the well-studied example of thermal and quantum tunneling in de Sitter space \cite{Coleman:1980aw,Brown:2007sd}. 
De Sitter space exhibits all of the regimes described above. For high enough Gibbons-Hawking temperature there is only the pure thermal instanton (called the `Hawking-Moss' in this case). For intermediate temperature there is both the Hawking-Moss instanton and the thermally-assisted quantum instanton (the `Coleman-De Luccia'), as well as intermediate solutions with extra negative modes \cite{Hackworth:2004xb}. For low enough temperature the Hawking-Moss itself has an extra negative mode and only the Coleman-De Luccia instanton contributes to the decay rate \cite{Jensen:1983ac}.

\subsection{Review of black strings and `caged' black holes}
\label{subsec:cagedblackholes}

In five-dimensional Kaluza-Klein spacetime, there are two possible static horizon topologies. There are black holes, for which the horizon topology is $S^3$; and there are black strings, for which the horizon topology is $S^2 \times S^1$. The phase diagram of these solutions has been extensively studied \cite{Kol:2004ww,Harmark:2005pp,LivingReviews,Harmark:2007md};  I now give a brief review. To characterize the black objects we will use their Hawking temperature, which is inversely proportional to their size: smaller means hotter.

\begin{figure}[h!] 
   \centering
   \includegraphics[width=5in]{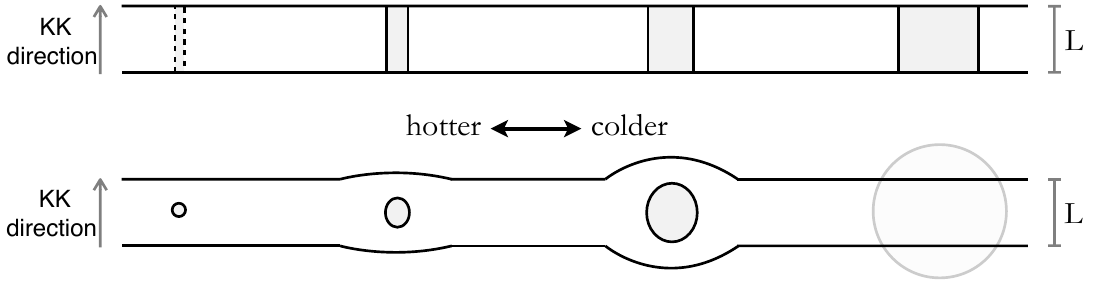} 
   \caption{Black strings (top) and `caged' black holes (bottom) as a function of temperature. The shaded region is the interior of the event horizon. At high temperatures the black hole is small and the black string is thin;  for $\beta < (1.75 \ldots)L$ the black string is so thin that it has a mechanical `Gregory-Laflamme' instability to become nonuniform.   At lower temperatures the black hole is larger and is elongated in the KK direction.  Eventually, for $\beta > (3.39 \ldots) L$, the would-be black hole is too large to be accommodated and there is no such solution.}
   \label{fig-caged}
\end{figure}

A  black string of uniform girth wrapped around an extra dimension of circumference $L$ and in thermal equilibrium with a heat bath of temperature $T \equiv \hbar/\beta$ has free energy
\begin{equation}
F(\textrm{black string}) \equiv m - T S  =  \frac{1}{16 \pi} \frac{\beta L}{G_5}.
\end{equation}

\begin{figure}[h!] 
   \centering
   \includegraphics[width=5in]{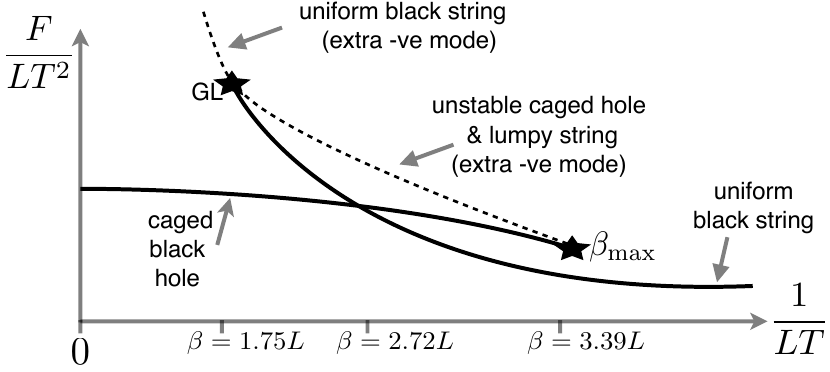} 
   \caption{Free energy of black strings and caged black holes as a function of their temperature. [This and all ensuing such diagrams are schematic---the precise curves can be straightforwardly plotted by adapting the numerical solution of 
   Fig.~5 of \cite{Headrick:2009pv}, but the result would be unsightly and unilluminating.] To accentuate the similarity to Fig.~\ref{fig-thermalandquantumQM} we have plotted $F$ in units of $LT^2$. This quantity would be independent of $T$ for an uncaged 5D black hole, but for $L< \infty$ the gravitational attraction to the image black holes bends the curve downwards \cite{Myers:1986rx}. Between the Gregory-Laflamme point and $\beta_{\textrm{max}}$ there is also a dynamically-unstable solution with an extra negative mode, discussed in Appendix~\ref{sec:mergerpoint}.} 
   \label{fig-freeenergyofholesandquantum}
\end{figure}

A five-dimensional black hole breaks the symmetry in the extra-dimensional direction and is therefore more complicated. If the black hole is much smaller than $L$ it doesn't `notice' the compactification, and so its free energy is  that of  five-dimensional Schwarzschild,
\begin{equation}
F(\textrm{black hole},L=\infty)  \equiv m - T S = \frac{1}{32 \pi} \frac{\beta^2}{G_5}.
\end{equation}
For $L < \infty$  the situation is more intricate. Harmark, Kol, \emph{et al.} analytically calculated the first-order \cite{Kol:2003if} and second-order \cite{Karasik:2004ds} corrections to the metric of these so-called `caged' black holes (we'll see their solution in Sec.~\ref{sec:blackholeinstanton}). By interrogating this metric we can extract the first- and second-order corrections to the free energy,
\begin{equation}
F(\textrm{`caged' black hole}, L < \infty) = \frac{1}{32 \pi} \frac{\beta^2}{G_5} \left( 1 - \frac{\beta^2}{16 L^2} + \frac{\beta^4}{128 L^4} +  \ldots  \right) . \label{eq:secondorderfreeenergy}
\end{equation}
The black hole's gravitational attraction to its images lowers the free energy at fixed asymptotic temperature.  For black holes of size comparable to the extra dimension the perturbative expansion breaks down and we will need the numerical results of     Headrick, Kitchen \& Wiseman \cite{Headrick:2009pv}. 

Figure~\ref{fig-freeenergyofholesandquantum} shows the phase diagram of black holes and black strings. There are four regimes:

\begin{enumerate}
\item[$\bar{4}.$] $(1.7524 \ldots) L >  \beta$: black hole only.

At high temperature the black string is so thin that it has a Gregory-Laflamme instability to become nonuniform \cite{Gregory:1993vy,Figueras:2011he}, and the only mechanically stable solution is the black hole.  

\item[$\bar{3}.$] $(2.72 \ldots) L >  \beta >   (1.7524 \ldots) L$: caged black hole beats black string.

Below the Gregory-Laflamme point the black string is mechanically stable, but at first still has higher free energy than the black hole. 

There is in addition a third solution to Einstein's equations, with an extra negative mode. 
This intermediate branch is mechanically unstable, and flows in one direction towards the uniform black string and in the other towards the caged black hole. This branch is composed of the nonuniform (or ``lumpy'') black string \cite{Gubser:2001ac} and the unstable caged black hole, and is discussed in Appendix~\ref{sec:mergerpoint}.

\item[$\bar{2}.$] $(3.39 \ldots)L >  \beta > (2.72 \ldots) L$: black string beats caged black hole. 

For $\beta = (2.72 \ldots) L$ the black string has the same free energy as the black hole. (This equality would have happened already at  $\beta = 2 L$ if not for the black hole's free-energy-lowering gravitational  attraction to its images.) 

\item[$\bar{1}.$] $ \beta > (3.39 \ldots) L$: black string only.

At $\beta_{\textrm{max}}$ the caged black hole meets the unstable caged black hole and annihilates \cite{Kol:2002xz}.  Below this temperature the would-be black hole is too large to be accommodated by the extra dimension and the only solution is the uniform black string.

\end{enumerate}

The phase diagram of quantum and thermal instantons resembles that of black holes and black strings, only with high and low temperatures switched. We will now see that this is not a coincidence.

\section{Black-string instanton} 
\label{sec:blackstringinstanton}
The Euclidean black-string instanton controls both the nucleation of black strings and the thermal nucleation of  bubbles of nothing\footnote{ This instanton actually mediates also a third transition, given by continuing not $w$ or $z$ but instead $\theta$. This gives the  quantum nucleation of a $2+1+2$-dimensional bubble of nothing in which the two extra dimensions, $w$ and $z$, are compactified on a torus and the $w$-direction pinches off. Since this is not an instability of hot  space with a single KK direction, it is beyond the remit of this paper.}. It is a vacuum solution of Einstein's equations that asymptotes to $S^1 \times S^1 \times R^3$ given by 
\begin{equation}
ds^2  =  (1 - \frac{R}{r} )d w^2 + \frac{dr^2 }{1 - \frac{R}{r} } + r^2 \left( d \theta^2 + \cos^2 \theta d \phi^2 \right)  + dz^2 . \label{eq:Euclideanblackstringmetric}
\end{equation}
To avoid a conical singularity at $r=R$, $w$ must be periodic under $w \rightarrow w + 4 \pi R$; $z$ may have any periodicity. The black-string instanton has a U(1) symmetry in the $w$-direction, a U(1) symmetry in the $z$-direction, and a SO(3) spherical symmetry. In Appendix \ref{appendix:EuclideanActions}, the Euclidean action is shown to be 
\begin{equation}
I_{E} = \frac{\pi R^2}{G_5}  \times \textrm{periodicity of }z.
\end{equation}

\begin{figure}[h!] 
   \centering
   \includegraphics[width=3in]{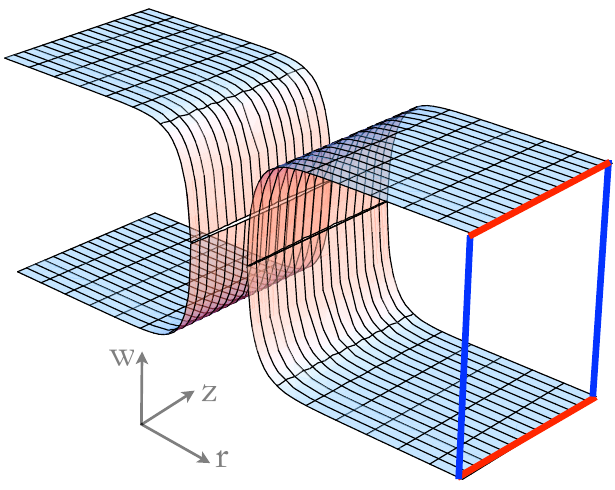} 
   \caption{The black-string instanton of Eq.~\ref{eq:Euclideanblackstringmetric}. The angular $S^2$ directions have been suppressed.}
   \label{fig-blackstringinstanton}
\end{figure}

The Euclidean black string always has at least one negative mode, associated with uniformly changing its radius. It can also have another one. As we saw in Sec.~\ref{subsec:cagedblackholes}, when the periodicity in the $z$-direction is more than $1.7524 $ times  the periodicity in the $w$-direction, there is another negative mode associated with the Euclidean black string becoming nonuniform.

\subsection{4D black holes/5D black strings} \label{subsec:5Dblackstrings}
Upon  analytic continuation $w \rightarrow it$,  the Euclidean instanton of Eq.~\ref{eq:Euclideanblackstringmetric} becomes
\begin{equation}
ds^2  =  -(1 - \frac{R}{r} )d t^2 + \frac{dr^2 }{1 - \frac{R}{r} } + r^2 \left( d \theta^2 + \cos^2 \theta d \phi^2 \right)  + dz^2 . \label{eq:Lorentzianblackstringmetric}
\end{equation}
The size of the extra dimension is the same everywhere. From the five-dimensional perspective this is a black string uniformly wrapped around the extra dimension; from the four-dimensional perspective this is a black hole. 

\begin{figure}[h!] 
   \centering
   \includegraphics[width=2.5in]{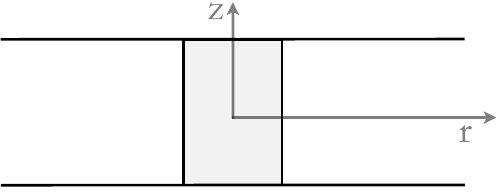} 
   \caption{The black string is given by a constant-$w$ slice through the instanton of Fig.~\ref{fig-blackstringinstanton}.} 
   \label{fig-bs}
\end{figure}

Since the $z$-direction is to be matched to the extra dimension and the $w$-direction (being the continuation of the time direction) is to be matched to the thermal circle, the required periodicity is
\begin{eqnarray}
z & \rightarrow & z + L \\
w & \rightarrow & w + \beta.
\end{eqnarray}
To avoid a conical singularity requires  $R = \beta /4 \pi$. 
%
%
%
%
The  rate to nucleate black strings is given by the instanton's Euclidean action
\begin{equation}
\Gamma \sim \exp \Bigl[ -\frac{ I_E }{\hbar} \Bigl] = \exp \Bigl[ -   \frac{1}{16 \pi} \frac{L \beta^2 }{G_5 \hbar }  \Bigl] = \exp  \Bigl[ -  \frac{1}{16 \pi} \frac{\beta^2 }{G_4 \hbar}  \Bigl].  \label{eq:actionblackstring}
\end{equation}
This expression agrees with that found in \cite{Gross:1982cv}.\\

The instanton gives the optimal route to traverse the barrier; we'll now  examine the barrier itself. 
The free energy of a black string of radius $R$ [and therefore of Hawking temperature $T_\textrm{Hawk} = \hbar/(4 \pi R)$] in a heat bath of temperature $T$  is 
\begin{equation}
F  \equiv m - TS  = \frac{L}{2 G_5} \left( {R -   \frac{2 \pi T R^2}{\hbar} } \right) .
\end{equation}
This function, plotted below, has a local minimum at $R=0$ and a global minimum as $R \rightarrow \infty$, but to get from one to the other---to go from  no string to  a huge string---involves traversing a barrier. A barrier may in principle be traversed either quantum mechanically or thermally, but this particular barrier is so broad that only the pure thermal route is locally optimal. (Since the barrier itself depends on $T$, we can be in the $T>T_{\textrm{\st{quant}}}$ regime for all $T$.) The only option is to go straight to the top. 
\begin{figure}[h!] 
   \centering
   \includegraphics[width=3.8in]{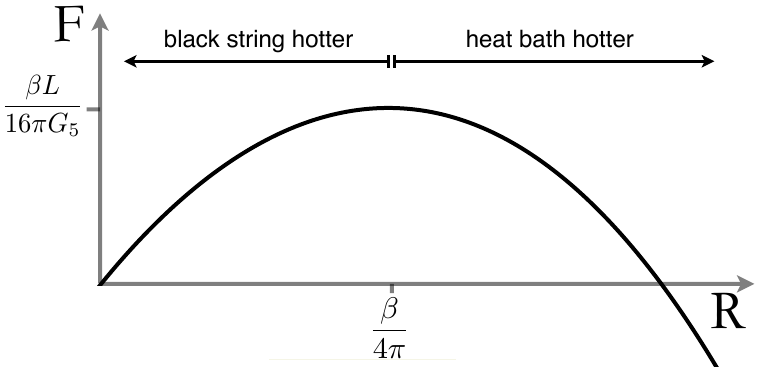} 
   \caption{The free energy of a black string, as a function of its radius $R$.} 
   \label{fig-FofBS}
\end{figure}

The top of the barrier is at $R = \beta / 4 \pi$, where the Hawking temperature of the black string is equal to the temperature of the heat bath, $T_{\textrm{Hawk}} = T$: the string is nucleated in thermal equilibrium with the radiation. 
But the equilibrium is an unstable one. If the string shrinks a little then it gets hotter (its specific heat is negative), loses more mass, gets hotter still, and is soon reclaimed by the heat bath. Conversely if the string grows a little then it cools and keeps on growing. The  negative mode of the Euclidean instanton has been continued to a thermodynamic instability of the nucleated string.

The exponentially most likely way to make a black string that lives forever is to make one just over the threshold for survival. Since no tunneling is required, only thermal fluctuation, the rate to make this plucky string is given by its Boltzmann suppression $\exp[-F/T]$. This rate is the same as that given by the instanton calculation since for solutions with a U(1) symmetry round the thermal circle the Euclidean action is related to the free energy by 
\begin{equation}
I_E = \frac{{\hbar} F}{T} = \beta m - \frac{S}{\hbar}. \label{eq:ActionIsFreeEnergy}
\end{equation}
The equivalence holds only for U(1)-symmetric thermal circles. For decays that feature quantum tunneling as well as thermal fluctuation (and therefore which break U(1) to Z$_2$) the action is not given by Eq.~\ref{eq:ActionIsFreeEnergy} and the free energy of the decay product does not calculate the decay rate.

For $\beta < (1.7524 \ldots) L$, the Euclidean instanton has in addition a second negative mode. This mode is continued to a mechanical Gregory-Laflamme instability of the black string. For higher temperatures, therefore,   the uniform black string is not a locally-optimal decay product.

\subsection{Thermal bubbles of nothing} 
Upon  analytic continuation $z \rightarrow it$,  the Euclidean instanton of Eq.~\ref{eq:Euclideanblackstringmetric} becomes
\begin{equation}
ds^2  =  (1 - \frac{R}{r} )d w^2 + \frac{dr^2 }{1 - \frac{R}{r} } + r^2 \left( d \theta^2 + \cos^2 \theta d \phi^2 \right)  - dt^2 . \label{eq:tBoNmetric}
\end{equation}
The size of the extra dimension is smaller for small $r$, eventually pinching off entirely at $r=R$; for $r<R$ there is no space and no time---there is literally nothing. Unlike the original bubble of nothing discovered by Witten (the `quantum' bubble of nothing in our vocabulary, see Eq.~\ref{eq:quantumBoNmetric}), 
this `thermal' bubble of nothing is static \cite{Sorkin:1983ns}.

\begin{figure}[h!] 
   \centering
   \includegraphics[width=2.5in]{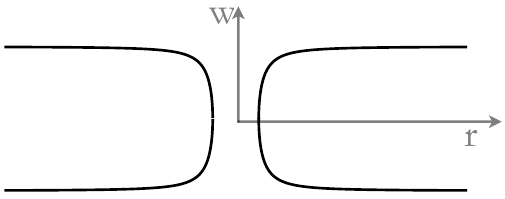} 
   \caption{The thermal bubble of nothing, given by a constant-$z$ slice through the instanton of Fig.~\ref{fig-blackstringinstanton}.} 
   \label{fig-tBoN}
\end{figure}

Since the $w$-direction is to be matched to the extra dimension and the $z$-direction is to be matched to the thermal circle, the required periodicity is
\begin{eqnarray}
w & \rightarrow & w + L \\
z & \rightarrow & z + \beta .
\end{eqnarray}
To avoid a conical singularity requires $R = L/4 \pi$. The rate to nucleate a thermal bubble of nothing is given by the instanton's Euclidean action
\begin{equation}
\Gamma \sim \exp \Bigl[ - \frac{I_E }{\hbar} \Bigl] = \exp \Bigl[ -   \frac{1}{16 \pi} \frac{L^2  }{G_5 T }  \Bigl] = \exp  \Bigl[ -  \frac{1}{16 \pi} \frac{L}{G_4 T}  \Bigl].\label{eq:thermalBoNaction}
\end{equation} 
 
 The barrier to nucleation is provided by the fact that small bubbles of nothing automatically seal up. 
Consider the minimum free energy for a spacetime that pinches off at a given area-radius $R$. This function goes from zero at  $R=0$ to minus infinity as $R \rightarrow \infty$ \cite{Brill:1989di}, but to get from one to the other---to go from no bubble to a huge bubble---involves traversing a barrier, and the thermal bubble of nothing sits at the top of this barrier.

The top of the barrier is a precarious place to be. If the radius shrinks slightly the bubble of nothing collapses under its own surface tension\footnote{The collapse will presumably give rise to a  black string, but since the mass of the static bubble of nothing is so small, the black string will itself have a Gregory-Laflamme instability and soon become nonuniform \cite{Corley:1994mc,Harmark:2007md}.}, if the radius grows slightly  the bubble wall accelerates outwards and  grows forever. The negative mode of the Euclidean instanton has been continued to a mechanical instability of the nucleated bubble.

Since no tunneling is required to make a thermal bubble of nothing, only thermal fluctuation, the decay rate is  given by the Boltzmann suppression of the decay product,  $\exp[-F/T]$. Said another way,  since there is a U(1) symmetry round the thermal circle, $I_E = \beta F$.

For $L < (1.7524 \ldots)  \beta$, the Euclidean instanton has a second negative mode. This mode indicates that pure thermal tunneling is no longer even locally the fastest path across the barrier (indicates that $T< T_{\textrm{\st{therm}}}$, in the language of Sec.~\ref{subsec:thermalandquantum}). 

\subsection{Black string \emph{vs}. thermal bubble of nothing}  \label{sec:stringVStBON}

The properties of the nucleated objects are (see Appendices~\ref{appendix:EuclideanActions} \& \ref{appendix:ADMmasses})
 \begin{center}
\begin{tabular}{c||c|c}
& black string & thermal BoN   \B \\
  \hline   \hline
ADM mass & $\frac{1}{8 \pi } \frac{\beta L}{G_5 } $ & $\frac{1}{16 \pi } \frac{L^2}{G_5 } $   \T\B \\
\hline entropy & $\frac{1}{16 \pi } \frac{\beta^2 L}{\hbar \, G_5 } $&  0  \T\B \\
\hline action & $\frac{1}{16 \pi } \frac{\beta^2 L}{G_5 } $  &  $\frac{1}{16 \pi } \frac{\beta L^2}{G_5 } $ \T\B 
\end{tabular}
\end{center}
The two Euclidean actions are related by the duality $\beta \leftrightarrow L$, which swaps the labels on the two Euclidean $S^1$s. At the self-dual point, $\beta = L$, the two  actions, and therefore the two free energies, are identical. But the duality applies only to Euclidean quantities, and does not extend to Lorentzian quantities like mass and entropy. The four-dimensional black hole is heavier than the thermal bubble of nothing (twice as heavy), but precisely compensates for having to borrow more from the heat bath with its large entropy.
%

\begin{figure}[htbp] 
   \centering
   \includegraphics[width=6in]{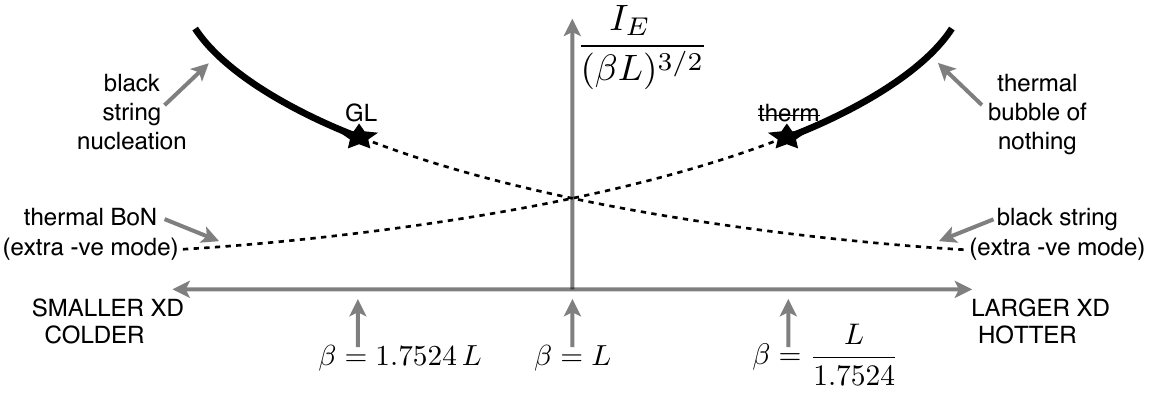} 
   \caption{The Euclidean action $I_E$ of the instantons that make a black string and a thermal bubble of nothing, in units of $ \beta^{3/2} L^{3/2}/G_5$  [schematic]. The tunneling rate is $\Gamma \sim \exp [ - I_E /\hbar]$, so smaller $I_E$ means faster tunneling. Above the Gregory-Laflamme temperature the string has an extra negative mode; below $T_{\textrm{\st{therm}}}$ the thermal tunneling instanton has an extra negative mode (and so isn't even locally the fastest way across the barrier). A $\beta \leftrightarrow L$ duality permutes the decays.}
      \label{fig-stringVSthermal}
\end{figure}

Figure~\ref{fig-stringVSthermal} compares the two rates.
At high temperature ($T L > \hbar$) it is faster to nucleate a black string; at low temperature ($ TL<\hbar$) it is faster to thermally nucleate a bubble of nothing. There is a $ \beta \leftrightarrow L$ duality that permutes the decays.

\section{Black-hole instanton}  \label{sec:blackholeinstanton}

The Euclidean black-hole instanton controls both the nucleation of 5D black holes and the quantum nucleation of  bubbles of nothing. It is a vacuum solution to Einstein's equations that asymptotes to $S^1 \times S^1 \times R^3$. The Euclidean black hole has a U(1) symmetry in the $w$-direction and an SO(3) spherical symmetry, but no  U(1) symmetry in the $z$-direction; the location of the Euclidean black hole in the $z$-direction breaks the symmetry down to Z$_2$. Because of the reduced symmetry, there is not a general analytic solution for the metric.

\begin{figure}[h!] 
   \centering
   \includegraphics[width=3in]{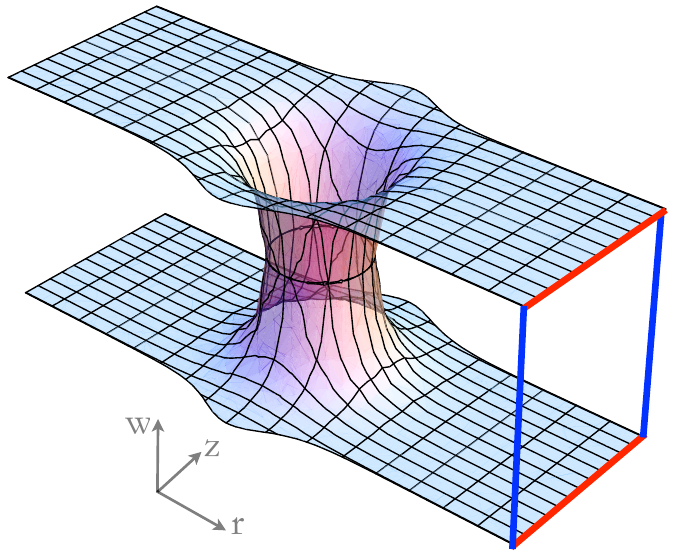} 
   \caption{The black-hole instanton. The angular  $S^2$ directions have been suppressed.} 
   \label{fig-blackholeinstanton}
\end{figure}

\subsubsection{Ignoring periodicity of $z$}
In the limit where the $z$-direction is so large as to be effectively uncompactified, the metric is the Euclidean 5D Schwarzschild solution 
\begin{equation}
ds^2 = (1 - \frac{R^2}{r^2} )d w^2 + \frac{dr^2 }{1 - \frac{R^2}{r^2} } + r^2 \left( d \theta^2 + \cos^2 \theta \left( d \phi^2 + \sin^2 \phi \, d \psi^2 \right)  \right) , \label{eq:Euclideanblackholemetric}
\end{equation}
where $z$ is defined cylindrically as $z = r \sin \theta$. To avoid a conical singularity at $r=R$, $w$ must be periodic under $w \rightarrow w + 2 \pi R$. 
The Euclidean action is derived in Appendix~\ref{appendix:EuclideanActions} to be 
\begin{equation}
I_{E} = \frac{1}{G_5} \frac{ \pi^2 R^3}{4} .
\end{equation}

\subsubsection{Including periodicity of $z$}

For Euclidean black holes that are  much smaller than the periodicity of the $z$-direction, a perturbative expression for the metric\footnote{To first order in $\epsilon$, when the asymptotic periodicity of $z$ is $\bar{L}$,  the close-in metric is 
\begin{eqnarray}
ds^2 &=& \left( 1 - \frac{ \epsilon \bar{L}^2}{\rho^2} - \frac{\pi^2 \epsilon }{12 \rho^2} \left[ 4 (1 - \sin^4  \theta ) (\rho^2 - \epsilon \bar{L}^2) + \epsilon \bar{L}^2 \right] \right)  dw^2 \nonumber \\
&&  \ \ \ \ + \frac{1 + \frac{\pi^2 \epsilon}{3} \left[ (3 \frac{\rho^2}{\epsilon \bar{L}^2} - 1) \sin ^4  \theta  - 1 \right] }{ 1 - \frac{ \epsilon \bar{L}^2}{\rho^2} - \frac{\pi^2 \epsilon }{12 \rho^2} \left[ 4 (1 - \sin^4  \theta ) (\rho^2 -  \epsilon \bar{L}^2) +  \epsilon \bar{L}^2 \right] } d \rho^2 - 2 \pi^2  \frac{\rho^3}{  \bar{L}^2} \sin^3  \theta  \cos  \theta  \,  d \rho \, d  \theta  \nonumber \\ 
&& \ \ \ \ \ \  \ \ \ + \rho^2 \left( 1 + \pi^2 \epsilon \left( \frac{\rho^2}{ \epsilon \bar{L}^2} - 1\right) \sin^2  \theta  \cos^2  \theta    \right)  d  \theta ^2 + \rho^2 \cos^2  \theta  d \Omega_2^2 , \label{eq:closetoblackhole}
\end{eqnarray}
which is a slightly prolate version of Eq.~\ref{eq:Euclideanblackholemetric}. This matches via $z \sim \rho \sin   \theta $ onto the far-out metric 
\begin{eqnarray}
ds^2 & =& (1 - \epsilon \frac{\pi \bar{L} }{r}  \frac{\sinh \frac{2 \pi r}{\bar{L}} }{\cosh \frac{2 \pi r}{\bar{L}}  - \cos \frac{2 \pi z}{\bar{L}}  } ) dw^2 + ( 1 - \epsilon \frac{\pi \bar{L}} {2r} \frac{\sinh \frac{2 \pi r}{\bar{L}} }{\cosh \frac{2 \pi r}{\bar{L}}  - \cos \frac{2 \pi z}{\bar{L}}  }  ) dr^2 \nonumber \\ 
& & \hspace{6cm} + r^2 d \Omega_2^2 + (1 - \epsilon \frac{ \pi \bar{L}}{2r} \frac{\sinh \frac{2 \pi r}{\bar{L}} }{\cosh \frac{2 \pi r}{\bar{L}}  - \cos \frac{2 \pi z}{\bar{L}}  }  ) dz^2 . \label{eq:farfromblackhole}
\end{eqnarray}
The value of $\epsilon$ is  set by requiring that there be no conical singularity at the pinch-off point $\rho = \sqrt{\epsilon} \bar{L} (1 + \frac{\pi^2 \epsilon}{24})$. The far-out metric makes manifest the Z$_2$ symmetry (both the $z=0$ and $z = \bar{L}/2$ fixed points) whereas the close-in region only overlaps with one of the fixed points (the $z = 0$ fixed point, equivalent to $\theta =0$).} is given by suitably analytically continuing the results of Harmark, Kol, \emph{et al.} \cite{Kol:2003if}. For Euclidean black holes that are comparable in size to the periodicity of the $z$-direction, we will use the numerical results of  Headrick, Kitchen \& Wiseman \cite{Headrick:2009pv}. Euclidean black holes of size much larger than the periodicity of the $z$-direction do not exist.  As we saw in Sec.~\ref{subsec:cagedblackholes}, when the periodicity of the $w$-direction is more than $3.39$ times the periodicity of the $z$-direction, the would-be Euclidean black hole is too large to be accommodated. Whenever it does exist, the Euclidean black hole has exactly one negative mode.

\subsection{5D black holes}

The 5D black hole is given by continuing the black-hole instanton $w \rightarrow it$. 
\begin{figure}[h!] 
   \centering
   \includegraphics[width=2.5in]{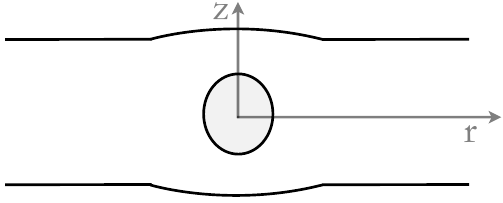} 
   \caption{The five-dimensional caged black hole, given by a constant-$w$ slice through the instanton of Fig.~\ref{fig-blackholeinstanton}.} 
   \label{fig-bh}
\end{figure}

\subsubsection{Ignoring periodicity ($L = \infty$)}
Upon  analytic continuation $w \rightarrow it$,  the Euclidean instanton of Eq.~\ref{eq:Euclideanblackholemetric} becomes
\begin{equation}
ds^2 = -(1 - \frac{R^2}{r^2} )d t^2 + \frac{dr^2 }{1 - \frac{R^2}{r^2} } + r^2 \left( d \theta^2 + \cos^2 \theta \left( d \phi^2 + \sin^2 \phi \, d \psi^2 \right)  \right) . 
\end{equation}
This is the $4+1$-dimensional Schwarzschild solution.
%
Since the $w$-direction is to be matched to the thermal circle, the required periodicity is
\begin{equation}
w \rightarrow w + \beta.
\end{equation}
To avoid a conical singularity requires $R= \beta/ 2 \pi$.
In uncompactified 5D Minkowski, the rate to nucleate a black hole is 
\begin{equation}
\Gamma \sim \exp \Bigl[ - \frac{I_E}{\hbar} \Bigl] = \exp \Bigl[ - \frac{1}{32 \pi } \frac{\beta^3}{G_5 \hbar} \Bigl] .
\end{equation}

\subsubsection{Including periodicity ($L < \infty$)} 
Upon  analytic continuation $w \rightarrow it$,  the Euclidean instanton of Eqs.~\ref{eq:closetoblackhole} \& \ref{eq:farfromblackhole} becomes a `caged' black hole, localized in the extra dimension. The required periodicity is 
\begin{eqnarray}
z & \rightarrow & z + L \\
w & \rightarrow & w + \beta.
\end{eqnarray}

Since the instanton has a U(1) symmetry around the thermal circle---which is to say since no quantum tunneling is required, only thermal fluctuation---the Euclidean action can be determined directly from the free energy of the decay product. We already calculated the free energy of a caged black hole in Eq.~\ref{eq:secondorderfreeenergy}, so using  $I_E = \beta F$ the nucleation rate is
\begin{equation}
\Gamma \sim \exp \Bigl[ - \frac{I_E}{\hbar}  \Bigl]  = \exp \left[ -  \frac{\beta^3 }{32 \pi G_5   \hbar} \left( 1 - \frac{\beta^2}{16 L^2 } +  \frac{\beta^4}{128 L^4 }  +\ldots  \right) \right] . \label{eq:peturbativecorrectionstoaction}
\end{equation}
The leading-order correction makes decay faster---the gravitational attraction between the hole and its images makes it easier to nucleate a five-dimensional black hole. 

Just as was true for the black string, the exponentially most likely way to make a caged black hole that lives forever is to make one just over the threshold for survival. The black hole is born in unstable thermal equilibrium with the heat bath, so the negative mode of the Euclidean instanton has been continued to a thermodynamic instability.

\subsection{Quantum bubbles of nothing}
The quantum bubble of nothing is given by continuing the black-hole instanton $z \rightarrow it$. 
\begin{figure}[h!] 
   \centering
   \includegraphics[width=2.5in]{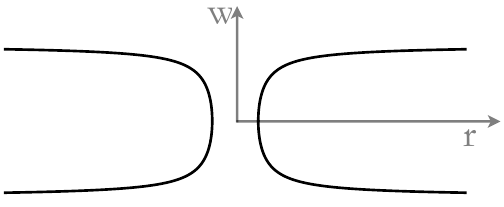} 
   \caption{A snapshot of the (thermally-assisted) quantum bubble of nothing at nucleation, given by a constant-$z$ slice through the instanton of Fig.~\ref{fig-blackholeinstanton}. At the moment of nucleation the bubble is at rest; it then expands, accelerating outwards and annihilating spacetime.} 
   \label{fig-qBoN}
\end{figure}

\subsubsection{Ignoring thermal assist $(\beta = \infty)$}
Upon  analytic continuation $z \rightarrow i t$ (given in the coordinates of Eq.~\ref{eq:Euclideanblackholemetric} by continuing $\theta \rightarrow i\hat{t} $, where $t= r \sinh \hat{t}$),  the Euclidean instanton becomes
\begin{equation}
ds^2 = (1 - \frac{R^2}{r^2} )d w^2 + \frac{dr^2 }{1 - \frac{R^2}{r^2} } + r^2 \left(- d \hat{t}^2 + \cosh^2 \hat{t} \left( d \phi^2 + \sin^2 \phi \, d \psi^2 \right)  \right) .  \label{eq:quantumBoNmetric}
\end{equation}
This is the `quantum' bubble of nothing. The size of the extra dimension shrinks for small $r$, eventually pinching off entirely at $r=R$; for $r<R$ there is no space and no time---there is literally nothing. Unlike the three  other Lorentzian spacetimes considered so far, this spacetime is not static. Instead the bubble expands,  a wall of annihilation that leaves nothing in its wake. 

Since the $w$-direction is to be matched to the extra dimension, the required periodicity is
\begin{equation}
w \rightarrow w + L .
\end{equation}
To avoid a conical singularity requires $R=L/2 \pi$. At zero temperature, 
the rate to nucleate a quantum bubble of nothing is
\begin{equation}
\Gamma \sim \exp \Bigl[ - \frac{I_{E}}{\hbar} \Bigl]  = \exp \Bigl[ -  \frac{1}{32 \pi} \frac{L^3}{G_5 \hbar} \Bigl]  =  \exp \Bigl[ - \frac{1}{32 \pi} \frac{ L^2}{G_4 \hbar } \Bigl]. \label{eq:WittenWrong}
\end{equation}
This exponent is a factor of two smaller\footnote{For independent confirmation that the answer quoted in  \cite{BoN} is out by a factor of two, see  \cite{Dowker:1995gb}. They calculate the rate for the quantum nucleation of a bubble of nothing in the presence of a magnetic field; taking the $B \rightarrow 0$ limit of their  rate (their Eq.~8) recovers my Eq.~\ref{eq:WittenWrong}.}
than that found by Witten \cite{BoN}.

\subsubsection{Including thermal assist ($\beta < \infty$)}

The Euclidean black-hole instanton does not have a U(1) symmetry in the $z$-direction. If we interpret the $z$-direction as the thermal circle, this mean that not all slices of constant Euclidean time are equivalent, and so it matters \emph{which} slice we continue. To get a real Lorentzian section we must continue one of the two fixed points of the Z$_2$ symmetry. These two choices give rise to two different Lorentzian spacetimes with two different interpretations:
\begin{itemize}
\item[-] continuing $z \rightarrow it$ gives the spacetime as it appears after the nucleation of a bubble of nothing. In the language of Sec.~\ref{subsec:thermalandquantum}, this is analogous to  $\bar{\bar{x}}$.

[This continuation surface straddles both the far-out coordinates of Eq.~\ref{eq:farfromblackhole} and the close-in coordinates of Eq.~\ref{eq:closetoblackhole}. For the close-in coordinates,  $z \rightarrow it$ is given by $ \theta   \rightarrow i \hat{t}$.]
\item[-] continuing $ z  \rightarrow \frac{\beta}{2} + i t$  gives the spacetime \emph{after} thermal fluctuation but \emph{before} quantum tunneling. In the language of Sec.~\ref{subsec:thermalandquantum}, this is analogous to  ${\bar{x}}$.

 [This continuation surface avoids the close-in region entirely.]
\end{itemize}  
The required periodicity is 
\begin{eqnarray}
w & \rightarrow & w + L \\
z & \rightarrow & z + \beta.
\end{eqnarray}

The rate to nucleate a thermally-assisted quantum bubble of nothing is given by acting with the duality $L \leftrightarrow \beta$ on Eq.~\ref{eq:peturbativecorrectionstoaction},
\begin{equation}
\Gamma \sim \exp \left[ - \frac{L^3}{32 \pi G_5  \hbar } \left( 1 - \frac{L^2 }{16  \beta^2} +  \frac{L^4 }{128 \beta^4 }  +  \ldots  \right) \right]  .
\end{equation}
The thermal assist speeds decay.  The instanton gives the local optimum of all paths across the barrier, including those that take different amounts of energy from the heat bath, so the effect of the thermal assist can only be to make the quantum decay faster. 

The other three processes considered in this paper proceeded by pure thermal fluctuation, with no quantum tunneling required, so the decay rate was given by the Boltzmann suppression of the decay product, $\exp[-F/T]$. The thermally-assisted-quantum-bubble-of-nothing decay features quantum tunneling as well as thermal fluctuation, and is therefore further suppressed over and above the Boltzmann suppression by a WKB tunneling factor: the absence of a U(1) symmetry round the thermal circle means that for the quantum bubble of nothing, $I_E > \beta F$.

\subsection{Black hole \emph{vs.} quantum bubble of nothing}
\begin{figure}[htbp] 
   \centering
   \includegraphics[width=6in]{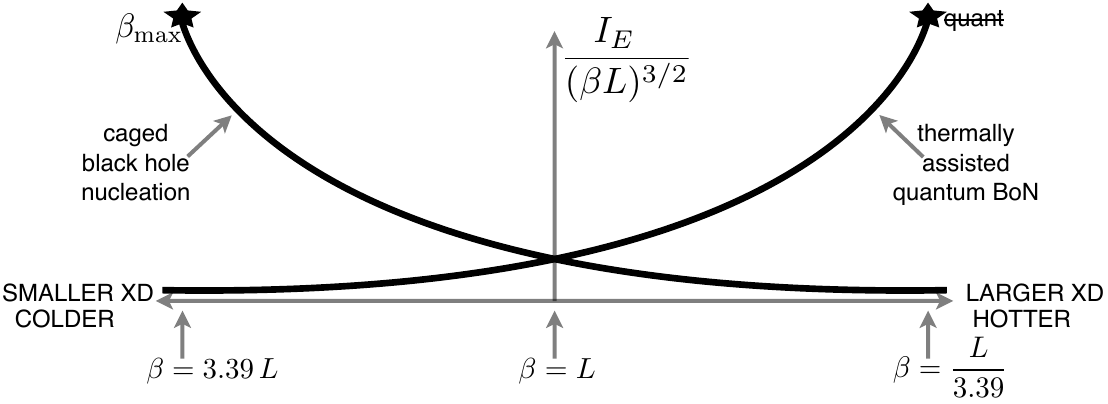} 
      \caption{The Euclidean action $I_E$ of the instantons that make a `caged' black hole and a quantum bubble of nothing, in units of $ \beta^{3/2} L^{3/2}/G_5$  [schematic]. The tunneling rate is $\Gamma \sim \exp [ - I_E/\hbar ]$, so smaller $I_E$ means faster tunneling. 
A $\beta \leftrightarrow L$ duality permutes the decays. Black holes only exist for temperatures above $T=\beta_\textrm{max}^{\, -1}$; the quantum bubble-of-nothing instanton only exist for temperatures below $T_{\textrm{\st{quant}}}$.}
   \label{fig-holeVSquantum}
\end{figure}

Figure~\ref{fig-holeVSquantum} compares the two rates.
At high temperature ($T L > \hbar$) it is faster to nucleate a black hole; at low temperature ($ TL<\hbar$) it is faster to  nucleate a quantum bubble of nothing. There is a $ \beta \leftrightarrow L$ duality that permutes the decays.  

\newpage

\section{Discussion}
\label{sec:grandcomparison}

If you are inside a black hole you cannot get out---the gravitational field is too strong. If you are outside a bubble of nothing you cannot get in---there  is no `in'.

It has long been known that bubbles of nothing may form  by quantum tunneling \cite{BoN}. We have seen that bubbles of nothing may also form by thermal fluctuation, or by a mixture of thermal fluctuation and quantum tunneling. Just as in quantum field theory, the thermal instanton has a U(1) symmetry around the thermal circle, whereas the quantum instanton has only a Z$_2$ symmetry. And just as in quantum field theory,  after nucleation the bubble bears the imprint of how it was made: the thermal-made bubble is born smaller than the quantum-made bubble ($R = \frac{L}{4 \pi}$ \emph{vs.} $R = \frac{L}{2 \pi}$); the thermal bubble has positive energy whereas the quantum bubble has zero energy; and the thermal bubble is in (unstable) static equilibrium whereas the quantum bubble, while born  at rest, immediately accelerates outwards.

\begin{figure}[htbp] 
   \centering
   \includegraphics[width=6in]{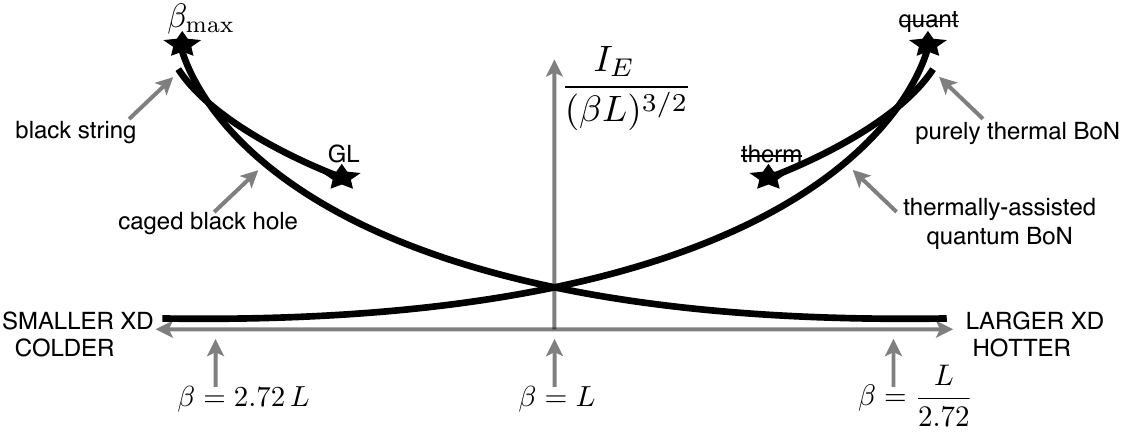} 
   \caption{The Euclidean action $I_E$ of the instantons that control each of the four non-perturbative instabilities of hot  KK space, in units of $ \beta^{3/2} L^{3/2}/G_5$  [schematic]. The tunneling rate is $\Gamma \sim \exp [ - I_E/\hbar ]$, so smaller $I_E$ means faster tunneling. A $\beta \leftrightarrow L$ duality permutes the decays. [Not shown: solutions with extra negative modes, such as the lumpy black string,  the black string past the Gregory-Laflamme point, or their corresponding duals.]} 
   \label{fig-altogether}
\end{figure}

Figure~\ref{fig-altogether} shows the decay rate for all four non-perturbative instabilities of hot  KK space. 
For low temperature $(TL<\hbar)$ the globally fastest decay is to make a bubble of nothing via a thermally-assisted quantum process. Indeed, for these temperatures that's the only locally-optimal way to make a bubble of nothing, since the thermal process has an extra negative mode. It is also possible to nucleate black holes and black strings, and even though these processes are exponentially subdominant at low temperatures, they lead to distinct endpoints, and so are independently interesting. Starting at the lowest temperature, the locally-optimal ways to make black objects go though our four familiar regimes:

\begin{enumerate}
\item[$\bar{1}.$] $ \beta > (3.39 \ldots) L$: black string only.

\item[$\bar{2}.$] $(3.39 \ldots)L >  \beta > (2.72 \ldots) L$: black string beats caged black hole. 

\item[$\bar{3}.$] $(2.72 \ldots) L >  \beta >   (1.7524 \ldots) L$: caged black hole beats black string.

\item[$\bar{4}.$] $(1.7524 \ldots) L >  \beta$: caged black hole only.
\end{enumerate}
At $ \beta = L$ there is an exchange of dominance between black holes and bubbles of nothing. For high temperature $(TL>\hbar)$ the globally fastest decay is to make a caged black hole. Indeed, for these temperatures that's the only locally-optimal way to make a black object, since the black string has an extra negative mode. It is also possible to nucleate bubbles of nothing, and even though these processes are exponentially subdominant at high temperatures, they lead to distinct endpoints, and so are independently interesting. As we keep raising the temperature, the locally-optimal ways to make a bubble of nothing go though our four regimes, the exchange now playing out in reverse:
\begin{enumerate}
\item[4.]$(1.7524 \ldots)   \beta > L$: thermally-assisted quantum bubble of nothing only.

\item[3.] $(2.72 \ldots)  \beta > L >   (1.7524 \ldots)  \beta$: thermally-assisted quantum beats pure thermal.

\item[2.] $(3.39 \ldots)  \beta > L > (2.72 \ldots)  \beta$: pure thermal beats thermally-assisted quantum.

\item[1.] $L > (3.39 \ldots)  \beta$: thermal bubble of nothing only. 

\end{enumerate}
At the highest temperatures, the fastest way to make anything is to make a black hole, and the fastest way to make nothing is to do so thermally. 
\pagebreak

We have seen that there is a duality that relates the nucleation of bubbles of nothing to the nucleation of black holes and black strings. The duality acts only on  Euclidean quantities, like the Euclidean action or the number of negative modes, and does not apply to Lorentzian quantities, like mass or entropy. The duality acts with the following dictionary: 

\vspace{1mm}
 \begin{center}
\begin{tabular}{c|c}
Bubbles of Nothing & Black Holes and Black Strings \B \\
  \hline   \hline
$\beta$ \& $L$   & $L$ \& $\beta$ \T\B \\
  \hline   \hline
thermal BoN  & black string \T\B \\
  \hline
quantum BoN   & black hole \T\B \\
\hline
thermally-assisted quantum BoN   & `caged' black hole \T\B \\
\hline
most-improbable BoN & unstable caged hole/nonuniform  string  \T\B \\  
    \hline   \hline
    mechanical instability & thermodynamic instability \T \\
    of thermal BoN & of black string in heat bath \B \\
    \hline
thermal `assist' of quantum BoN & attraction between image black holes \T \B \\
    \hline  \hline
quantum BoN disappears  & black hole ceases to exist \T \\
at $T_{\textrm{\st{quant}}}$  & at $\beta_\textrm{max}$ \\
 $L = (3.39 \ldots) \beta$ & $\beta = (3.39 \ldots)L$ \B \\
\hline
exchange of dominance between  & exchange of dominance between \T \\
 thermal and  quantum BoN & black string and black hole  \\
 $L =  (2.72 \ldots )\beta $ & $\beta = (2.72 \ldots) L$ \B \\
\hline 
thermal BoN no longer locally  & Gregory-Laflamme instability \T \\
optimal path across barrier &  of black string \\
 $L = (1.7524 \ldots)  \beta$ & $\beta = (1.7524 \ldots )L$ \B \\
 \hline \hline
\end{tabular}
\end{center}
 
 \newpage
 
The nucleation of a bubble of nothing changes the spatial topology. Before nucleation, the topology is $R^3 \times S^1$; after nucleation, space has a hole and the topology is $R^2 \times S^2$. While there is not believed to be anything wrong with topology changes in quantum gravity \cite{Wheeler:1957mu}, in classical general relativity topology changes are governed by restrictive no-go theorems \cite{Horowitz:1990qb}.  
It is therefore worth asking whether the thermal bubble of nothing should be thought of as an example of a classical topology change. 

What we usually mean by a process being `classical' is that the rate stays nonzero as we take the classical limit, $\hbar \rightarrow 0$. However, the ultraviolet catastrophe means that in thermal field theory there is no obvious unique  classical limit, and the answer will depend  on \emph{which} classical limit we take. Including factors of $\hbar$, the rate to nucleate a thermal bubble of nothing is
\begin{equation}
\Gamma( \textrm{thermal BoN}) \sim  \exp \Bigl[ -  \frac{1}{16 \pi}  \frac{ \beta}{\hbar} \frac{ L^2}{G_5}  \Bigl]  \sim   \exp \Bigl[ -   \frac{1}{16 \pi}  \frac{1 }{\hbar^{3/4} \rho^{1/4} }   \frac{L^2}{G_5}  \Bigl] \sim \exp \Bigl[ - \frac{1}{16 \pi} \frac{1}{T} \frac{L^2}{G_5}  \Bigl] , \label{eq:differentclassicallimits}
 \end{equation}
 where $\beta \equiv \hbar /T$ is the thermal wavelength of the radiation and $\rho \equiv T/ \beta^3$ is proportional to its energy density. Equation~\ref{eq:differentclassicallimits} displays three different `classical' limits one might be tempted to consider; they all  take $\hbar \rightarrow 0$, but they differ in what happens to $T$. The first limit keeps  $\beta$ fixed: the wavelength of the thermal radiation stays constant (this is the limit advocated by \cite{Gross:1982cv}). In this limit  the decay rate scales as $\exp[-1/\hbar]$ so the topology change is `quantum'. The second limit keeps  $\rho$ fixed: the energy density of the thermal radiation stays constant. Here too the topology change is `quantum', though with a weaker power of $\hbar$. The final limit keeps  $T$ fixed: the mean energy of a photon stays constant. In this limit, there are no factors of $\hbar$ in the decay rate, and the topology change is `classical'. However, I would caution that  this `classical' limit is rather eccentric: the wavelength is going to zero, the energy density is going to infinity, the Jeans instability is becoming short-wavelength, and not only is the topology change classical, but so too is the photo-electric effect. \\

Wrapping the Euclidean black string around the extra dimension describes the nucleation of a Lorentzian black string; wrapping it around the thermal circle describes the thermal nucleation of a bubble of nothing. But the  asymptotic Euclidean $T^2$ has more cycles than just these two. For every coprime $\{ n_1 ,n_2 \}$ there is a cycle that goes $n_1$ times round the $z$-direction and $n_2$ times round the $w$-direction. Since the $\{ n_1 , n_2 \}$-cycle of a  rectangle is related by an SL$(2,\mathbb{Z})$ transformation to the $\{1,0\}$-cycle of a parallelogram, I will defer discussion of these extra solutions until I turn on an angle between the two S$^1$s, which is to say until I turn on a KK chemical potential \cite{me:chemicalpotential}. It will still remain true that when the chemical potential is zero the  four non-perturbative decays described in this paper exponentially dominate the decay rate. \\

In higher dimensions the phase diagram of black holes and black strings is known to differ from the five-dimensional case studied in this paper. There are new phenomena, like stable nonuniform black strings \cite{Sorkin:2004qq}, and a more intricate pattern of appearances and disappearances. It would be interesting to repeat the analysis of this paper in higher dimensions, and see whether thermal bubbles of nothing can ever be the globally fastest decay. To do this analysis would require extending the numerical work of \cite{Headrick:2009pv,Sorkin:2003ka} to higher dimensions, but thankfully with the same cohomogeneity. In Appendix \ref{appendix:GLF} I show that one thing that we can say without any new numerics is that in  eight or more dimensions the two halves of Fig.~\ref{fig-altogether} meet, in the sense that at the self-dual point $\beta = L$ all four instantons still have only one negative mode and therefore are  all still in play. \\

This paper has considered the decay of hot  KK space in the \emph{canonical} ensemble. What if we moved to the  \emph{microcanonical} ensemble, by enclosing our system in a box and demanding that it has fixed energy, not fixed temperature? 
For the largest boxes, this makes no difference---the heat bath is so huge that the temperature does not change during nucleation. However, as we will see in Appendix~\ref{appendix:microcanonical}, for smaller boxes an appreciable fraction of the heat bath's energy condenses into the decay product, and the temperature of the radiation appreciably drops. This punishes thermal decays relative to quantum decays, and in so doing breaks the duality between black holes and bubbles of nothing. \\

The entropy of a thermal-made bubble of nothing was calculated in Sec.~\ref{sec:stringVStBON}, and found to be zero. This is as it should be. If unperturbed, the thermal bubble of nothing just sits there forever; it can be circumambulated at leisure and gives rise to no Lorentzian horizons. Not so the quantum-made bubble of nothing. The quantum bubble of nothing expands so vigorously \cite{Aharony:2002cx} that it casually separates antipodal observers---the hole in spacetime grows faster than it can be walked around---and one might imagine that those Lorentzian horizons have an associated entropy. Unfortunately the U(1) symmetry around the thermal circle that made it easy to calculate the entropy of a thermal bubble of nothing is absent from the quantum case. With a colleague, Xi Dong, I am  adapting the method of \cite{Lewkowycz:2013nqa} to calculate the entropy of a quantum bubble of nothing. \\

The florid final paragraphs of a certain kind of theoretical physics paper often quote Robert Frost's  meditation on whether the world will end with fire or with ice (and, this being poetry, on how those two competing possibilities make Frost feel). An echo of this choice appears in the two fates of hot KK space outlined above: whether to be consumed by a black hole that cannot be sated, or to be annihilated by a bubble of nothing that cannot be stopped.  Frost, the great dualist,  would no doubt have found it fitting that these two fates are really just two sides of the same coin, that there is a high-temperature/low-temperature duality relating black hole to bubble of nothing, consumption to annihilation, and fire to ice.

\section*{Acknowledgements}
Thanks to Alex Dahlen, Sean Hartnoll, Matthew Headrick, Shamit Kachru, Loganayagam Ramalingam, Jorge Santos, Steve Shenker,  and I-Sheng Yang. Valedictions to Alex Dahlen, a good friend and the finest collaborator since P\'etain; Silicon Valley's gain is physics' loss, and my loss too. Thank you to the Aspen Center for Physics and the NSF Grant \#1066293 for hospitality during part of the writing of this paper.

\appendix

\section{Perturbative instabilities}
\label{appendix:perturbative}

Hot KK space is a dangerous place. The Jeans instability makes overdensities grow; the gravitational backreaction of the thermal radiation and of the Casimir energy makes both extended and compact dimensions change size. For $\beta \sim L$ the perturbative length scale is
\begin{equation}
\textrm{Hubble length} 
 \sim  \textrm{Jeans Length}
\sim  \frac{ \beta^2}{\ell_4} \sim \frac{ L^2}{\ell_4} \label{eq:JeansLength}.
\end{equation}
The existence of perturbative instabilities threatens to swamp the non-perturbative instabilities considered in this paper. What can prevent this swamping is the separation of scales. The characteristic length scales of the non-perturbative decays are 
\begin{eqnarray}
\textrm{size of  critical black hole/string} & \sim &  \beta \label{eq:BHvsJeans} \\
\textrm{size of  critical bubble of nothing} & \sim &  L, 
\end{eqnarray}
which are much shorter than the perturbative decays (for  $\beta \gg \ell_5$  \& $L  \gg \ell_5$), roughly speaking   because non-linear gravity is much stronger than linear gravity. This means that by cutting off our theory far out beyond $\beta$ and $L$ but before the Jeans/Hubble scale, we can eliminate the perturbative instabilities while retaining and isolating the non-perturbative ones. 

There's  an important intermediate scale, given by the linear size of the region whose radiation must be harvested in order to make the decay product. For $\beta \sim L$ this is given by 
\begin{equation}
\textrm{mass of critical black hole/black string/BoN } \sim \textrm{ mass of radiation in box of  size }  \frac{ \beta^{5/3}}{\ell_4^{2/3}}.
\end{equation}
This is the scale at which the nucleated object, while still only accounting for a tiny fraction of the total volume, starts to account for an appreciable fraction of the total energy. (For a  fixed total energy, this is the AdS-length at which large Hawking-Page black holes first become thermodynamically stable \cite{Hawking:1982dh}.)   Since the nucleation  changes the temperature of the heat bath,  for boxes of this size we must move to the microcanonical ensemble.



\section{Microcanonical ensemble}
\label{appendix:microcanonical}

If the temperature of the heat bath changes during the nucleation process, what $T$ should be used in $\exp[-\Delta E/T]$, and what $\beta$ should be used to set the periodicity of the Euclidean instanton? The partition function approach  does not reveal the answer, but the rather more direct approach embodied in Eq.~\ref{eq:GammaMixedQM} does. The Boltzmann factor is replaced by the change in entropy of the heat bath
\begin{equation}
\exp\Bigl[- \frac{\Delta E}{T} \Bigl] \ \  \rightarrow \  \ \exp \biggl[ - \int_{0}^{{\Delta E}} \frac{dE}{T[E]} \biggl] ,
\end{equation}
and since the mixed-strategy transition involves \emph{first} fluctuating and \emph{then} tunneling,  the  periodicity of Euclidean time is  not the inverse temperature before the transition but rather the inverse temperature after the transition.

\section{Euclidean actions} 
\label{appendix:EuclideanActions}

The $n+1$-dimensional Euclidean Schwarzschild metric is 
\begin{equation}
ds^2 = \left(1 - \frac{R^{n-2}}{r^{n-2}} \right) dw^2  + \left(1 -  \frac{R^{n-2}}{r^{n-2}}\right)^{-1} d r^2 + r^2 d \Omega^2_{n-1}.
\end{equation}
Near $r=R$ the angular directions  peel off as a constant-sized $n-1$-sphere, leaving an effective two-dimensional cone. To leading order 
\begin{equation}
ds^2 =  \left(\frac{n-2}{2R} \right)^2 y^2 dw^2+ dy^2 + R^2 d \Omega_{n-1}^2,
\end{equation} 
where 
\begin{equation}
y \equiv \frac{2 R}{n-2}   \sqrt{1 -  \frac{R^{n-2}}{r^{n-2}} }.
\end{equation}
For there to be no conical singularity, $w$ must be periodic under
\begin{equation}
w \rightarrow w + 2 \pi \frac{2 R}{n-2}, \label{eq:conicalgeneralanswer}
\end{equation}
which for $n=3$ comes to $4 \pi R$ and for $n=4$ comes to $2 \pi R$. \\

\noindent In $n+1$ dimensions  the action of a Euclidean Schwarzschild black hole is 
\begin{equation}
I_E = -\int_{\mathcal{M}} \frac{1}{16 \pi G} \mathcal{R} \sqrt{g}  - \int_{\partial \mathcal{M}} \frac{1}{8 \pi G} K \sqrt{h} .
\end{equation}
The Ricci curvature is identically zero for our Ricci-flat vacuum solutions, so the only contribution is the boundary term. There is no boundary at $r=R$, the metric is smooth there, and the only boundary is at $r \rightarrow \infty$. The extrinsic curvature is defined as 
\begin{equation}
K_{\mu \nu} \equiv \nabla_{\mu} n_{\nu} - n_{\mu} n^{\rho} \nabla_{\rho} n_{\nu},
\end{equation}
where $n^{\mu}$ is the normalized normal 
\begin{eqnarray}
n^{\mu}  & = &  \{0, \chi,0, \ldots,  0 \} \\
 n_{\mu} & = &  \{0, \chi^{-1},0, \ldots, 0 \} ,
\end{eqnarray}
and $\chi \equiv \left(1 - \frac{R^{n-2}}{r^{n-2}}\right)^{\frac{1}{2}}$. It is follows that 
\begin{equation}
K_{\mu \nu} = n_{\mu} \Gamma_{r \nu}^{r} - \Gamma_{\mu \nu}^r n_r.
\end{equation}
We have $K_{rr} =0$ as required. Since $\Gamma_{r \nu}^{r}=0$ (for $\nu \neq r$) and since $\Gamma^{r}_{\mu \nu} = -\frac{1}{2} g^{rr} \partial_r g_{\mu \nu}$ (for $\mu, \nu \neq r$), we have  
\begin{equation}
K_{\mu \nu} = \frac{1}{2} \chi \partial_r g_{\mu \nu}  \ \   \textrm{ for } \ \mu, \nu \neq r. \label{eq:extrinsicofgeneralform}
\end{equation}
The extrinsic curvature scalar is then 
\begin{equation}
K = g^{ww}K_{ww} + (n-1) g^{\theta \theta}K_{\theta \theta}  =  \frac{1}{2 \chi} \partial_r \chi^2 + (n-1)\frac{\chi}{2  r^2} \partial_r r^2 =  \frac{(n-2)R^{n-2}}{2 \chi r^{n-1}}  + \frac{(n-1) \chi}{r}    . \label{eq:extrinsiccurvature}
\end{equation}
(Notice that $g^{tt}K_{tt}$ recovers the proper acceleration of a constant-$r$ trajectory.) If we took the same surface and embedded it in flat space ($R^4 \times S^1$) then the extrinsic curvature would be 
\begin{equation}
K_0 = \lim_{R \rightarrow 0} K = \frac{n-1}{r} .
\end{equation}
The action is given by 
\begin{equation}
I_E =  -\lim_{r \rightarrow \infty} \frac{1}{8 \pi G}  \int_{\partial \mathcal{M}}  (K-K_0) \sqrt{h} =  \frac{1}{n-2} \frac{{ R^{n-1} \Omega_{n-1}}}{4 G}  =  \frac{1}{n-2} S.
\end{equation}
For $n=3$ this 
reduces to the result of Hawking and Gibbons' \cite{Gibbons:1976ue}. We use the $n=3$ result in Sec.~\ref{sec:blackstringinstanton} and the $n=4$ result in Sec.~\ref{sec:blackholeinstanton}.  \\

\noindent For Euclidean solutions such as these that have a U(1) symmetry around the thermal circle, the Euclidean action is the  Helmholtz free energy divided by the temperature
\begin{equation}
I_E = \beta F = \beta m - \frac{S}{\hbar}.
\end{equation}
In $n+1$ dimensions the Schwarzschild mass scales as $R^{n-2}$, whereas the entropy scales as $R^{n-1}$ so 
\begin{equation}
S \sim m^\frac{n-1}{n-2} \ \ \  \rightarrow \ \ \ 
\frac{1}{T} \equiv \frac{\partial S}{\partial m}  = \frac{n-1}{n-2} \frac{S}{m} .
\end{equation}
In terms of the entropy, we therefore have
\begin{eqnarray}
\textrm{Boltzmann suppression factor}  =   \frac{m}{T} &=&   \frac{n-1}{n-2} S  \\
\textrm{entropy of black hole}    & =& S  \\
 \textrm{decay exponent} =  I_E/\hbar & = &  \frac{1}{n-2} S.
\end{eqnarray}
The Euclidean action is equal to the entropy only when $n=3$.

\section{ADM masses}
\label{appendix:ADMmasses}

Consider a metric whose asymptotic behavior is given to first order in $r^{-1}$ by  
\begin{equation}
ds^2 = - \left( 1 - \frac{2 G_4 M_t}{r} \right) dt^2 + \frac{dr^2}{1 - 2 G_4 M_r /r}  + r^2 d \Omega_2^2 + \left( 1 - \frac{2 G_4 M_z}{r} \right) dz^2 . 
\end{equation}
If this is to correspond to a vacuum solution, $\mathcal{R} = 0$, then necessarily
\begin{equation}
M_r = M_t + M_z. \label{eq:vacuumcondition}
\end{equation}
(This condition follows either from direct calculation or from the combination of linearity, knowing the $M_z = 0$ answer, and knowing that there is a symmetry between $M_t$ and $M_z$ if we continue to Euclidean spacetime.) We will now show that  the ADM mass for this metric is
\begin{equation}
m = M_r  - \frac{1}{2} M_z = M_t  + \frac{1}{2} M_z = \frac{1}{2} \left( M_r + M_t \right) . \label{eq:massasfunctionofschwarzschildparameters}
\end{equation}
(Notice  that even though there is a $t \leftrightarrow z$ symmetry in the vacuum condition, Eq.~\ref{eq:vacuumcondition}, there is no such symmetry in the expression for the ADM mass, since the ADM mass is a `Lorentzian' quantity that cares about \emph{which} Lorentzian continuation you take. This asymmetry underlies the fact that  for $ \beta = L$  there is still a difference between the mass of the critical black string and the mass of the thermal bubble of nothing, and also a difference between the mass of the critical black hole and the  mass of the quantum bubble of nothing.) 

The ADM formalism asks us to transform to asymptotically Cartesian coordinates, a request we can accommodate with $r^2 = \vec{x} \cdot \vec{x}$. In these coordinates the metric is 
\begin{eqnarray}
dr^2 &=& \left(\frac{\vec{x} \cdot d \vec{x}}{r} \right)^2 \\
r^2 d \Omega^2 &=& d \vec{x} \cdot d \vec{x} - dr^2 
\end{eqnarray}
\begin{equation}
ds^2 = - \left( 1 - \frac{2 G M_t}{r} \right) dt^2 + \left[ \left(1 - \frac{2 G M_r}{r} \right)^{-1} - 1 \right]  \left(\frac{\vec{x} \cdot d \vec{x}}{r} \right)^2 + d \vec{x}^2 + \left( 1 - \frac{2 G M_z}{r} \right) dz^2 . 
\end{equation}
The leading-order perturbations of this metric from flat KK space are
\begin{eqnarray}
h_{ij}& = & 2 GM_r \frac{x_i x_j}{r^3} \\
\partial_k h_{ij} & = &  2 GM_r  \frac{1}{r^3} \left[ \delta_{ik} x_j + \delta_{jk} x_i - 3 \frac{x_i x_j x_k}{r^2} \right] \\
\partial_k h_{zz} & = & \frac{2 G M_z}{r^3} x_k.
\end{eqnarray}
The original formula for the ADM energy was written down by ADM \cite{Arnowitt:1962hi}, the  generalization to cases with a compact extra dimension was given in Eq.~2.7 of \cite{Deser:1988fc} as, in our notation, 
\begin{eqnarray}
m & = & \lim_{r \rightarrow \infty} \sum_{i,j} \frac{1}{16 \pi G_5} (2 \pi L) (4 \pi r^2 \frac{x_k}{r} ) \left[ \partial_j h_{kj} - \partial_k h_{jj} - \partial_k h_{zz} \right] \label{eq:sumovertheseindices} \\
&= & \frac{r}{4 G_4}  \left[ \frac{2 G M_r}{r^3} 2 \vec{x}^2  - \frac{2 G M_z}{r^3} \vec{x}^2  \right] \\
& = & M_r - \frac{1}{2} M_z , \label{eq:ADManswer}
\end{eqnarray}
recovering Eq.~\ref{eq:massasfunctionofschwarzschildparameters}. (The same answer may be reached by integrating out the extra dimension and converting to four-dimensional Einstein frame; the Weyl factor rescales the four-dimensional metric  by $\sqrt{1 - 2M_z/r}$, and leads to Eq.~\ref{eq:ADManswer}.) \\

For completeness, now let's calculate not just the ADM energy but also the ADM momentum and tension: the full ADM stress tensor. We will use Eq.~3.13 of \cite{Hovdebo:2006jy}, 
\begin{equation}
T_{a b} = \frac{4 \pi r^2}{16 \pi G} n^{i} \left[ \eta_{ab} \left( \partial_i h^{c}_{\ c} + \partial_i h^{j}_{\ j} - \partial_j h^{j}_{\ i} \right) - \partial_i h_{ab} \right] ,
\end{equation}
where $\{ a, b, c \}$ range over $\{ z, t \}$ and $\{ i, j , k\}$ range over the other directions. 

 For the  black string, the ADM stress tensor comes to
\begin{equation}
T_{a b} =  M_r  \left( \begin{array}{cc}
1 &  0 \\
0   & - \frac{1}{2}   \label{eq:ADMwiseTofboostedstring}
\end{array} \right) .
\end{equation}
The ``$-$'' in the ``$-\frac{1}{2}$" indicates that the black string is under tension; the ``$\frac{1}{2}$'' in the  ``$-\frac{1}{2}$" shows that this tension is smaller than the mass-per-unit-length. 

For the thermal bubble of nothing, the ADM stress tensor comes to
\begin{equation}
T_{a b} =  M_r \left( \begin{array}{cc}
\frac{1}{2}  & 0\\
0   & -1  \label{eq:ADMwiseTofboostedBoN}
\end{array} \right) .
\end{equation}
The tension is large---twice as large as the mass-per-unit-length---so  large that the thermal bubble of nothing violates the ADM version of the Null Energy Condition. (The point-by-point Null Energy Condition, by contrast, is trivially saturated since we have a vacuum solution.) The equation of state parameter is given by $w = -2$, which could have been anticipated either from knowing that the mass scales super-linearly with the length   $m \sim L^2/G_5$, or by taking the dual of the black string parameter $w \rightarrow 1/w$. As with any stress tensor that violates the NEC, by longitudinal boosting we can reach a frame in which the energy is arbitrarily negative. 

\section{Gregory-Laflamme in higher dimensions} \label{appendix:GLF}
In five spacetime dimensions, the black string has a Gregory-Laflamme instability whenever  $\beta \leq \beta_{\textrm{GL}} = (1.7524 \ldots )L$. Since $1.7524$ is larger than $1$, this means that at the self-dual point, $\beta = L$, both the black string and the thermal bubble of nothing have an extra negative mode. (Pictorially, the lines in Fig.~\ref{fig-stringVSthermal} only meet after they have first become dashed.) 

In $n+1\geq8$ dimensions this is no longer true. Combining Eq.~\ref{eq:conicalgeneralanswer} and the numerical results reported in Table 1 of 
\cite{Figueras:2011he} and Table 1 of 
\cite{Kol:2004pn}, in $n+1$ spacetime dimensions $\beta_{\textrm{GL}}$ is 
\begin{table}[h]
\begin{center}
\begin{tabular}{c||c|c|c|c|c|c|c}
$n+1$ & 5 & 6 & 7 & 8 & 9 &10 & 100   \T \B \\
\hline
$\beta{_\textrm{GL} }$& 1.7524 L & 1.2689 L & 1.0539 L & 0.9243 L & 0.8349 L & 0.7696 L & 0.203 L  \T \B 
\end{tabular}
\end{center}
\vspace{-5mm}
\caption{The critical inverse temperature  for the  Gregory-Laflamme instability in $n+1$ spacetime dimensions.}
\label{table:GL}
\end{table}

\newpage

\section{Merger point}
\label{sec:mergerpoint}

In our discussion of the phase diagram of black strings and caged black holes, summarized in Fig.~\ref{fig-freeenergyofholesandquantum}, there was an inessential feature we omitted: the merger point. Let's discuss it now. 

\begin{figure}[htbp] 
   \centering
   \includegraphics[width=4.4in]{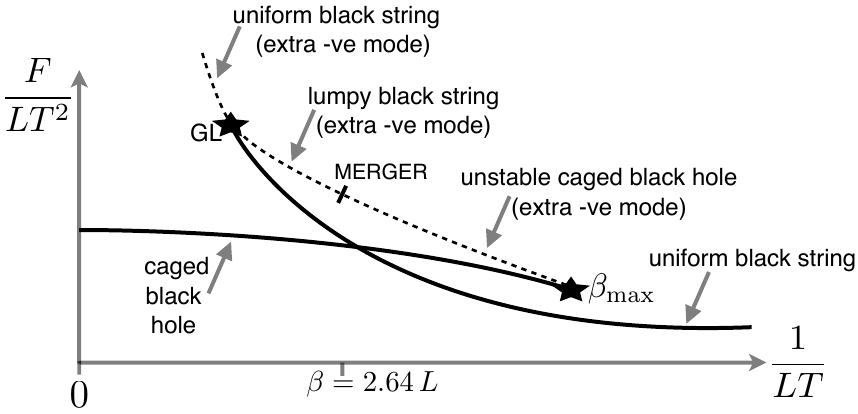} 
   \caption{The merger point lies on the unstable branch, and divides unstable lumpy black strings from unstable caged black holes. The merger point has been numerically calculated \cite{Headrick:2009pv} to lie at $\beta = (2.64 \ldots) L$.}
   \label{fig-mergerpoint}
\end{figure}

The merger point appears on the unstable branch and separates unstable caged black holes from lumpy black strings, as shown in Fig.~\ref{fig-mergerpoint}.  The lumpy black string is nonuniform,  with maximum girth at $z=0$ and minimum girth at $z= L/2$. As first envisaged by Kol \cite{Kol:2002xz}, and later numerically corroborated by \cite{Headrick:2009pv}, as the temperature falls the lumpy black string becomes lumpier and lumpier, until eventually its waist at $z=L/2$ pinches to zero and it becomes topologically a black hole. The merger thus changes the topology, but does not change the symmetry (both branches have SO(3)$\times$U(1)$\times$Z$_2$ symmetry), and does not change the number of negative modes (both branches have two negative modes and are mechanically unstable).

The merger point has a different interpretation on the bubble-of-nothing side of the duality. Both the lumpy black string and the unstable caged black hole are dual to `pessimal' tunneling solutions---tunneling solutions with an extra negative mode (which do not, therefore, correspond to locally-optimal decays). Both produce bubbles of nothing via a thermally-assisted quantum process, which means that there is first a thermal fluctuation to some spatial slice (the analogue of  $\bar{x}$ in Fig.~\ref{fig-halfwayupthenacross}) and then a quantum tunneling from there to a supercritical bubble of nothing (the analogue of $\bar{\bar{x}}$ in Fig.~\ref{fig-halfwayupthenacross}); the supercritical bubble then classically expands. The merger point tells us about  ``$\bar{x}$'', the configuration \emph{after} thermal fluctuation but \emph{before} quantum tunneling. The tunneling solution dual to the lumpy black string first thermally fluctuates to a small subcritical bubble of nothing, and then quantum tunnels to a large supercritical bubble of nothing:  $\bar{x}$  already features a small bubble of nothing. The tunneling solution dual to the unstable caged black hole first thermally fluctuates to a spatial slice with an extra dimension that, while smaller than in the vacuum, does not pinch off: $\bar{x}$ has no bubble of nothing at all.

\end{document}